\documentclass[11pt]{article}
\usepackage[top=1in,bottom=1in,left=1in,right=1in]{geometry}

\usepackage{amsmath}
\usepackage{amssymb}
\usepackage{graphicx}
\usepackage{listings}
\usepackage{enumitem}
\usepackage{hyperref}
\usepackage{algorithm, algpseudocode}

\usepackage{float}
\usepackage{bm}
\usepackage{subfig}
\usepackage{amsmath,amssymb,mathrsfs}
\usepackage{bbm}
\usepackage{url}
\usepackage{wrapfig}
\usepackage{multirow}
\usepackage[normalem]{ulem}
\usepackage{booktabs}

\usepackage{etoolbox}
\usepackage{url}
\usepackage[most]{tcolorbox}
\usepackage{xcolor}

\newcommand{\qedwhite}{\hfill \ensuremath{\Box}}

\newcommand{\lp}{\left(}
\newcommand{\rp}{\right)}
\newcommand{\lb}{\left[}
\newcommand{\rb}{\right]}
\newcommand{\lbp}{\left\{}
\newcommand{\rbp}{\right\}}
\newcommand{\lba}{\left\lvert}
\newcommand{\rba}{\right\rvert}

\newcommand{\mv}{\middle\vert}
\newcommand{\ul}{\underline}

\newcommand{\mcal}{\mathcal}

\newcommand{\mb}{\mathbf}
\newcommand{\bbm}{\mathbbm}
\newcommand{\mbb}{\mathbb}
\newcommand{\msf}{\mathsf}
\newcommand{\la}{\leftarrow}
\newcommand{\ra}{\rightarrow}

\newcommand{\eqDef}{\triangleq}
\newcommand{\diid}{\overset{\text{i.i.d.}}{\sim}}

\newcommand{\E}{\mathbb{E}}
\newcommand{\Var}{\mathsf{Var}}

\newcommand{\dra}{\overset{d}{\ra}}
\newcommand{\pra}{\overset{q}{\ra}}
\usepackage{enumitem }

\usepackage[round]{natbib} 
\newtheorem{theorem}{Theorem}[section]
\newtheorem{proposition}[theorem]{Proposition}
\newtheorem{lemma}[theorem]{Lemma}
\newtheorem{corollary}[theorem]{Corollary}
\newtheorem{definition}[theorem]{Definition}
\newtheorem{assumption}[theorem]{Assumption}
\newtheorem{remark}[theorem]{Remark}

\renewcommand{\paragraph}[1]{\noindent{\textbf{#1}}}
\newcommand{\footremember}[2]{%
    \footnote{#2}
    \newcounter{#1}
    \setcounter{#1}{\value{footnote}}%
}
\newcommand{\footrecall}[1]{%
    \footnotemark[\value{#1}]%
} 

\title{\textbf{Federated Experiment Design under Distributed Differential Privacy}}
\author{Wei-Ning Chen\footremember{stanford}{Stanford University}\footremember{meta}{Meta} \and Graham Cormode\footrecall{meta}\footremember{warwick}{University of Warwick} \and Akash Bharadwaj\footrecall{meta}\footnote{ByteDance} 
\and Peter Romov\footrecall{meta} \and Ayfer \"Ozg\"ur\footrecall{stanford}}
\date{} 

\begin{document}
\maketitle

\begin{abstract}
Experiment design has a rich history dating back over a century and has found many critical applications across various fields since then. 
The use and collection of users' data in experiments often involve sensitive personal information, so additional measures to protect individual privacy are required during data collection, storage, and usage. 
In this work, we focus on the rigorous protection of users' privacy (under the notion of differential privacy (DP)) while minimizing the trust toward service providers. 
Specifically, we consider the estimation of the average treatment effect (ATE) under DP, while only allowing the analyst to collect population-level statistics via secure aggregation, a distributed protocol enabling a service provider to aggregate information without accessing individual data. 
Although a vital component in modern A/B testing workflows, private distributed experimentation has not previously been studied.
To achieve DP, we design local privatization mechanisms that are compatible with secure aggregation and analyze the utility, in terms of the width of confidence intervals, both asymptotically and non-asymptotically. 
We show how these mechanisms can be scaled up to handle the very large number of participants commonly found in practice.
In addition, when introducing DP noise, it is imperative to cleverly split privacy budgets to estimate both the mean and variance of the outcomes and carefully calibrate the confidence intervals according to the DP noise. 
Last, we present comprehensive experimental evaluations of our proposed schemes and show the privacy-utility trade-offs in experiment design.
\end{abstract}

\allowdisplaybreaks
\section{Introduction}\label{sec:introduction}
Experimental design has a long history, tracing back to the early 1920s in the agricultural domain \citep{fisher1936design}, where statisticians used mathematical tools to design and analyze experiments. 
Since then, experimental design has found many applications, e.g., in chemistry, manufacturing, pharmaceuticals, and technology, etc. 
It not only enables the comparison of specific design alternatives, but also facilitates the production of generalizable knowledge to inform strategic decision-making.
When designing experiments to estimate or test the effect of a treatment (for example, a tech company launching a new feature in an existing product), a standard procedure is to divide participants into test and control groups, introduce changes (``the treatment'') to the test group, and collect feedback or outcomes from both groups to conduct further statistical analysis. 
When the test assignment is properly randomized and the estimators or tests for the outcomes are designed adequately, the analyst can infer the treatment effect and make decisions accordingly.

However, in many modern applications, such as pharmaceutical and online experimental designs, experimentation usually involves participants' private data, raising additional concerns about privacy and security. 
Thus, when designing and conducting experiments involving sensitive personal information, additional safeguards are desirable to protect it.

One way to enforce rigorous privacy for experiments is by restricting the final tests or estimators used to be differentially private (DP) \citep{dwork2006calibrating}. 
In brief, DP defines a formal notion of privacy that quantifies the amount of information leakage of an algorithm. 
It ensures that the output of a (randomized) algorithm $\mcal{A}$ does not depend strongly on the contribution of any one individual. 
To achieve DP, a standard approach is to add carefully calibrated noise to the test statistics (e.g., the Laplace or Gaussian mechanisms \citep{dwork2006calibrating, dwork2014algorithmic}) and only using the perturbed results in downstream tasks. 
This approach is usually referred to as ``Central DP'', since an analyst collects all the experimental data centrally before sanitizing the test statistics.
While Central DP schemes control the view of downstream tasks and are relatively straightforward to design, the analyst stores and processes all the raw users' data in the clear.
This not only requires the experiment participants to trust the analyst, but could make it challenging to comply with regulations on the storage of certain forms of personal data. 

To address the above issues, an alternative approach is to aggregate test data in a ``secure'' way so that only necessary population-level statistics are collected and that analysts can never see individual data.
Secure aggregation can be achieved by secure hardware or cryptographic multiparty computation (MPC) \citep{bencompleteness,damgaard2012multiparty} and is the focus of ``federated learning and analytics'' \citep{kairouz2019advances}. 
Secure aggregation alone does not provide any formal differential privacy guarantees.
To ensure DP, participants can locally randomize their data so that the securely aggregated outcome satisfies the standard DP requirement~\citep{dwork2006our}. 
This is referred to as Distributed DP (in contrast to Central DP) and is growing in prominence thanks to recent progress in practical aggregation protocols~\citep{bonawitz2016practical, bell2020secure}. 
With secure aggregation and Distributed DP, one can minimize the level of trust in the data analysts and service providers.

In this work, we focus on experimental design with Distributed DP. Specifically, we consider estimating and testing the average treatment effect (ATE), subject to DP and secure aggregation constraints. 
In our framework, to construct private protocols, we make use of a secure aggregation primitive that we refer to as SecAgg, which can be instantiated by \citet{bonawitz2016practical, bell2020secure}.

Our contributions are as follows:
\begin{itemize}
    \item 

    We present a framework that achieves a $(1-\alpha)$-confidence interval (CI) and a level-$\alpha$ test while ensuring distributed DP (defined in Section~\ref{sec:formulation}).  We analyze the width of private confidence intervals and provide asymptotic and non-asymptotic guarantees. Our non-asymptotic bounds are based on a version of empirical Bernstein inequality, which guides how to allocate privacy budgets in estimating mean and variance.

   \item 

    We incorporate the Poisson-binomial mechanism (PBM) \citep{chen2022poisson} in our framework as the local randomizer, which offers several advantages, including unbiased estimation, efficient memory (or communication) usage, and bounded sensitivities, letting downstream parties develop privatization mechanisms.

    \item

    To use PBM for experimental design, we develop an improved privacy accounting tool based on a novel bound on the R\'enyi divergence. This enhancement greatly enhances efficiency in large sample scenarios. When the objective is to obtain CIs instead of point estimators, we must collect second-moment information such as sample variance. We show, via SecAgg and DP, that this can be done by judiciously allocating privacy budgets for estimating sample mean and variance.
   \item 

    Last, our experimental study quantifies the trade-offs between privacy and utility. 
\end{itemize}

\subsection{Related Works}\label{sec:related_works}
\paragraph{Private causal inference and testing.}
The design of experiments to identify causal relations and average treatment effects is crucial in various domains \citep{imbens2015causal}; 
when experiments involve sensitive data, additional privacy protection is needed such as differential privacy (DP).
\citet{d2015differential} proposes DP mechanisms for summary statistics in causal inference, and \citet{lee2019privacy, niu2022differentially, ohnishi2023locally} consider estimating conditional average treatment effects (CATE) and propose private estimation of inverse propensity scores. 
These works default to a Central DP setting where a central data curator collects and privatizes test statistics, while \citet{ohnishi2023locally} explore Local DP without a trusted curator. 
In contrast, we address the experimental design problem using Distributed DP via secure aggregation as a better compromise between privacy and security.
Our experiment design problem is related to private hypothesis testing, which performs two-sample tests under DP when potential outcomes come from an unknown distribution. 
Previous work on two-sample tests has primarily focused on either Central DP \citep{rogers2017new, cai2017priv, raj2020differentially} or Local DP \citep{raj2020differentially}. 
This work is the first to consider experimentation under Distributed DP with secure aggregation. 
We also analyze the distribution-free setting, where no distributional assumptions are imposed on potential outcomes.

\paragraph{Private mean estimation.}  
The mechanisms in this paper are based on the difference-in-mean estimator, which relies on private mean estimation as a sub-routine. 
Differentially private mean estimation has been extensively studied under Central DP \citep{dwork2006calibrating, dwork2014algorithmic, balle2018improving, agarwal2018cpsgd, biswas2020coinpress} or Local DP~\citep{duchi2013local, bhowmick2018protection, chen2020breaking, feldman2021lossless}. 
In addition to obtaining a point estimator for the mean, it is also desirable to obtain a $(1-\alpha)$-confidence interval (CI) for a level-$\alpha$ test. 
Existing methods either privately estimate both sample means and variances separately \citep{du2020differentially, karwa2017finite, d2015differential} or use a private bootstrap \citep{brawner2018bootstrap}.
Our approach resembles the former but is compatible with secure aggregation and does not require a central data curator.  In addition, all of the previous methods, to our knowledge, study the \emph{asymptotic} CIs, while in this work, we also characterize the non-asymptotic coverage guarantees with finite $n$. The only exception that also considers non-asymptotic bounds is the recent work \citet{waudby2023nonparametric}. However, \citet{waudby2023nonparametric} considers a Local DP setting, so the analyst can directly estimate the mean and variance based on the locally private samples.

\paragraph{Secure aggregation and distributed DP.} 
Our methods aggregate test data using secure aggregation protocols (specifically, single-server aggregation) to achieve distributed DP without introducing bias. 
Single-server secure aggregation is performed via additive masking over a finite group \citep{bonawitz2016practical, bell2020secure}. 
However, to achieve provable privacy guarantees, secure aggregation is insufficient as the sum of local model updates may still leak sensitive information \citep{melis2019exploiting, song2019auditing, carlini2019secret, shokri2017membership}. 
For DP, participants have to privatize their raw data with local noise before secure aggregation \citep{dwork2006our}. 
This local noise has to be compatible with the secure aggregation protocol; candidate solutions include \citep{agarwal2018cpsgd, kairouz2021distributed, agarwal2021skellam, chen2022poisson}. 
Here, we aim to provide privacy guarantees in the form of  R\'enyi DP \citep{mironov2017renyi} because it allows for tracking the end-to-end privacy loss tightly.
We distinguish our Distributed DP model from Local DP  \citep{kasiviswanathan2011can, evfimievski2004privacy, warner1965randomized}, where data is perturbed on the client-side before the server collects it in the clear. 
Although simpler to implement, Local DP naturally suffers from poor privacy-utility trade-offs, as much more noise is introduced in total \citep{kasiviswanathan2011can, duchi2013local}. 
\section{Problem Setup and Preliminaries}\label{sec:formulation}
We formulate the experiment design problem via the \textit{Neyman-Roubin causal model}. 
When a service provider considers introducing a new feature to the public, it initiates a test phase by selecting a small group of users. 
This group is randomly divided into two: a test group where users are exposed to the new feature (referred to as the treatment), and a control group where users do not have access to the feature. 
The service provider collects responses from both groups, assesses the effects of the treatment, and, based on the evaluation, makes a decision regarding the launch of the new feature.

Formally, we define the experiment design problem as follows: for each test unit (``user'') $i \in [n]$, we introduce the randomized treatment assignment variable $T_i \in \{c, t\}$ (for the \ul{c}ontrol and \ul{t}est group, respectively), which indicates whether user $i$ receives the treatment or not. Additionally, we consider the potential outcomes $y_i(t), y_i(c) \in \mcal{Y}$ for user $i$ when receiving or not receiving the treatment, respectively. 
For a test unit $i$, the service provider can only observe one of its potential outcomes: $X_i \eqDef y_i(T_i)$. The quantity of interest is the sample average treatment effect (SATE): 
$$ \Delta_{\msf{s}}(\mb{y}) \eqDef \frac{1}{n}\sum_{i=1}^n y_i(t) - y_i(c). $$
Notice that under Neymann's original framework, the potential outcomes 
$$ \mb{y} \eqDef \lbp (y_i(t), y_i(c)) \mv i=1,...,n\rbp $$
are deterministic; only the treatment variable $T_i$'s are randomized. However, we can also impose distributional assumptions on the potential outcomes, i.e.,
$ y_i(c) \diid P_c \text{ and }  y_i(t)\diid P_t, $
and the quantity of interest is the population average treatment effect (PATE):
$$ \Delta_{\msf{p}}(P_c, P_t) \eqDef \E_{Y(t)\sim P_t, \, Y(c)\sim P_c}\lb Y(t) - Y(c)\rb.$$

Our goal is to test if $\Delta_{\msf{s}} > 0$ (or $\Delta_{\msf{p}} > 0$) at a given confidence level $\alpha$, which is equivalent to construct $(1-\alpha)$ confidence intervals of $\Delta_{\msf{s}}$ (or $\Delta_{\msf{p}}$).

\subsection{Secure aggregation and DP}
When the service provider has access to all the observable data, it can estimate $\Delta_{\msf{s}}$ via standard causal inference in statistics, social, and biomedical sciences (see, for instance, \citet{imbens2015causal}), compute sample variances of $y_i(c)$'s and $y_i(t)$'s, and construct confidence intervals accordingly.
However, when the samples $X_i$ are treated as sensitive, they should be aggregated securely so that only necessary information is revealed to the service providers. 

\paragraph{Secure aggregation.} Recently, distributed protocols based on multi-party computation (MPC), such as secure aggregation (SecAgg, \citet{bonawitz2016practical}), have emerged as powerful tools for securely aggregating population-level information from a group of users. 
Specifically, SecAgg enables a single server to compute the population sum and, consequently, the average of local variables while ensuring that no additional information, apart from the sum, is disclosed to the server or other participating entities. 
These properties make SecAgg well-suited for aggregating experiment results from users to estimate or test Average Treatment Effects (ATE). 
This is because test statistics used for ATE estimation can often be expressed as a function of the average of users' potential outcomes. 
However, when applying SecAgg in experiment design, it is important to note that SecAgg typically operates on a finite field, like most cryptographic MPC protocols. 
Thus, each outcome $X_i$ needs to be appropriately pre-processed (e.g., discretized) and mapped into a finite field.

\paragraph{Differential privacy.} Secure aggregation alone does not provide any provable privacy guarantees. 
Sensitive information may still be revealed from the aggregated population statistics, causing potential privacy leakage. 
To address this issue, differential privacy (DP, \citet{dwork2006calibrating}) has been adopted as the gold standard that ensures individual information is not leaked. Specifically, it requires the ATE estimator (or a CI of ATE) released by the service provider to meet the following guarantee:
\begin{definition}[Differential privacy]
We say an ATE estimator $\hat{\Delta}\lp X^n\rp$ is $(\varepsilon, \delta)$-DP, if for any two possible outcome sets $\mb{y} \eqDef \{(y_i(c), y_i(t)) | i= 1,...,n\}$ and $\mb{y}' \eqDef \{(y_i(c), y_i(t)) | i= 2,...,n\}\cup \{(y'_1(0), y'_1(1))\}$ differing in one user, we have
$$
\Pr\lbp  \hat{\Delta}\lp X^n \mv \mb{y}\rp \in \mcal{S} \rbp \leq e^\varepsilon\Pr\lbp \hat{\Delta}\lp X^n \mv \mb{y}'\rp \in \mcal{S} \rbp + \delta,$$
for any measurable set $\mcal{S}$.
\end{definition}
A common approach to achieve DP is adding properly calibrated noise (such as Gaussian noise with appropriate variance) to standard (non-private) ATE estimators. 
However, this requires users to trust the service provider as the server can see the unprivatized aggregated information.
To address this issue, one can instead \emph{locally} perturb individual outcome $X_i$ before secure aggregation via a local randomizer $\mcal{M}(X_i)$. 
When the local noise mechanism $\mcal{M}$ is designed in a way that the sum $\sum_i \mcal{M}(X_i)$ satisfies DP, i.e.,
\begin{equation}\label{eq:ddp}
    \Pr\lbp  \sum_i \mcal{M}(X_i) \in \mcal{S} \mv \mb{y}\rbp \leq e^\varepsilon\Pr\lbp  \sum_i \mcal{M}(X_i) \in \mcal{S} \mv \mb{y}'\rbp  + \delta,
\end{equation} 
and when $\mcal{M}(X_i)$'s are aggregated securely, one can ensure DP \emph{even if the service provider is not trusted}. The idea of combining secure MPC with local noise dates back to \citet{dwork2006our} and has been used extensively in private federated learning and analytics \citep{kairouz2021distributed, agarwal2018cpsgd, agarwal2021skellam}. The main challenge is that the local noise has to be discretized and compatible with secure aggregation; i.e., $\mcal{M}$ has to map $X_i$ into a space $\mcal{Z}$ (a finite field, e.g., the integers modulo a prime $p$) for SecAgg to work in.

In addition to the above $(\varepsilon, \delta)$-DP, we also use the following R\'enyi DP definition, which allows simpler and tighter privacy composition.
\begin{definition}[R\'enyi differential privacy]
We say an ATE estimator $\hat{\Delta}\lp X^n\rp$ is $(\alpha, \varepsilon(\alpha))$-DP, if for any two neighboring sets of possible outcomes $\mb{y}$ and $\mb{y}'$ that differ in one user, it holds that
\begin{align*}
    &D_\alpha\lp  \hat{\Delta}\lp X^n \mv \mb{y}\rp \middle\Vert \hat{\Delta}\lp X^n \mv \mb{y}'\rp \rp
    \eqDef \frac{1}{\alpha-1}\log \E_{X\sim \hat{\Delta}\lp X^n \mv \mb{y}\rp}\lb \lp \frac{f_{ \hat{\Delta}\lp X^n \mv \mb{y}\rp}(X)}{f_{ \hat{\Delta}\lp X^n \mv \mb{y}'\rp}(X)} \rp^\alpha \rb \leq \varepsilon(\alpha).
\end{align*} 
\end{definition}

Similarly, for a local randomizer $\mcal{M}:\mcal{X}\ra \mcal{Z}$, we can define the following distributed R\'enyi DP.
\begin{definition}[Distributed Renyi DP]\label{def:distributed_rdp}
A local randomizer $\mcal{M}$ is $(\alpha, \varepsilon(\alpha))$-DP, if, for any two neighboring outcome sets $\mb{y}$ and $\mb{y}'$ differing in one user:
$$  D_\alpha\lp  \sum_i \mcal{M}(X_i|\mb{y}) \middle\Vert \sum_i \mcal{M}(X_i|\mb{y}') \rp \leq \varepsilon(\alpha).$$
\end{definition}

\section{A Distributed DP Framework}\label{sec:main}
Our objective is to construct a $(1-\alpha)$-confidence interval for SATE and PATE (which can then be used to design a level-$\alpha$ test) while adhering to the distributed differential privacy (DP) constraint mentioned in equation \eqref{eq:ddp}. 
In Algorithm~\ref{alg:SATE_ddp_general}, we presented a general framework for causal inference using secure aggregation and distributed DP.

\begin{algorithm}
   \caption{ATE Estimation with Distributed DP}
   \label{alg:SATE_ddp_general}
\begin{algorithmic}
    \State {\bfseries Input:} treatment variables $T_1,...,T_n \in \{c, t\}$, outcomes $(y_1(T_1),...,y_n(T_n))$, randomizers $\mcal{M}_1, \mcal{M}_2$, privacy budgets $\varepsilon_1$ and $\varepsilon_2$,  $ \msf{obj} \in \lbp \text{`SATE'}, \text{`PATE'} \rbp$.
    \State {\bfseries Output:} an $(1-\alpha)$-CI for ATE.
    \\

    \Statex \(\triangleright\) \textbf{Local Randomization}
    \For{each user $i$}
    \State Obtains the observable outcome $X_i \eqDef y_i(T_i)$.\;
    \State Computes $\mcal{M}(X_i)$, and $\mcal{M}(X_i^2)$.\;
    \EndFor
    
    \Statex \(\triangleright\) \textbf{Aggregation}
    \State Server securely aggregates 
    $$ \begin{cases}
    \sum_{i\in S_t} \mcal{M}_1(X_i, n_t),\, \sum_{i \in S_t} \mcal{M}_2(X_i^2, n_t);\\ 
    \sum_{i\in S_c} \mcal{M}_1(X_i, n_c),\, \sum_{i\in S_c} \mcal{M}_2(X_i^2, n_c),
    \end{cases}$$
    where $S_c \eqDef \{ i: T_i = c \}$ and $S_t \eqDef \{ i: T_i = t \}$. \;
    \\
    \Statex \(\triangleright\) \textbf{Estimation}
    \State Estimates sample means and variances: 
    $$ \hat{\mu}_c\lp \sum_{i\in S_c} \mcal{M}_1(X_i, n_c) \rp \text{ and }\hat{\mu}_t \lp \sum_{i\in S_t} \mcal{M}_1(X_i, n_t) \rp; $$ 
    {\small
    $$ \hat{s}^2_c\lp \sum_{i \in S_c} \mcal{M}_2(X_i^2, n_c), \hat{\mu}_c \rp \text{ and }  \hat{s}^2_t\lp \sum_{i \in S_t} \mcal{M}_2(X_i^2, n_t), \hat{\mu}_t \rp.$$\;
    }
    \State Computes the diff-in-mean estimator
    $ \hat{\Delta} \eqDef \hat{\mu}_t - \hat{\mu}_c. $
    \State Computes the variance calibration term $\sigma_\msf{pr}^2 \lp \varepsilon, n_c, n_t\rp$ according to \eqref{eq:calibration}.
    \If {$\msf{objective}$ is `SATE'}
    \State Set $\hat{\sigma}^2_\msf{s} \eqDef \frac{n_cn_t}{n}\lp \frac{\sqrt{\hat{s}^2_t}}{n_t} + \frac{\sqrt{\hat{s}^2_c}}{n_c}\rp^2.$
    \State {\bfseries Return:} $\hat{\Delta}_\msf{s}\pm z_{1-\alpha/2}\cdot\lp \hat{\sigma}_\msf{s} + \sigma_{pr} \rp$.
    \EndIf
    \If {$\msf{objective}$ is `PATE'}
    \State Set $\hat{\sigma}^2_\msf{p} \eqDef \frac{\hat{s}^2_t}{n_t} + \frac{\hat{s}^2_c}{n_c}.$
   
    \State {\bfseries Return:} $\hat{\Delta}_\msf{p}\pm z_{1-\alpha/2}\cdot  \lp \hat{\sigma}_\msf{p} + \sigma_\msf{pr} \rp.$
    \EndIf
\end{algorithmic}
\end{algorithm}

In this framework, the server securely aggregates necessary information from the control and test groups separately, along with local randomizers $\mcal{M}_1$ and $\mcal{M}_2$. 
These randomizers satisfy the distributed DP conditions defined in Definition~\ref{def:distributed_rdp} and map individual observable outcomes $X_i$ and their second moments $X_i^2$ to the finite field on which secure aggregation operates. 
Specifically, we have:
\[\begin{cases}
\mcal{M}_1: \mcal{X} \times [n] \ra \mcal{Z};\\
\mcal{M}_2: \mcal{X}_2 \times [n] \ra \mcal{Z},
\end{cases}\]
where we use $\mcal{X}_2\eqDef \{ x^2 | x \in \mcal{X}\}$ to denote the collection of all possible second moments of the samples.
In the above notation, we allow the local randomizers to take the size of the control (or test) group, denoted as $n_c\eqDef\sum_{i=1}^n \left(1-\bbm{1}_{\{T_i = c \}}\right)$ (or $n_t \eqDef n-n_c$), as an input. 
This enables the local randomizers to calibrate the noise level based on the group size. 
As in Neyman-Robin's potential outcome framework, the test assignment variables $(T_1,...,T_n)$ follow a uniform distribution across all sequences containing `c' $n_c$ times.
We consider bounded observations and without loss of generality, we assume the outcome domain $\mcal{X}$ is centered at $0$:
\begin{assumption}\label{assumption:bdd_x}
    Let $\mcal{X} = [-R, R]$ (so we must have $\mcal{X}_2 = [0, R^2]$).
\end{assumption}
After receiving the aggregated information, the server constructs unbiased estimators for the sample means and variances of each group. 
The difference-in-means estimator is then used to estimate the ATEs. 
The second-moment information is used for variance estimation, which is needed for confidence intervals.

\subsection{Privacy of Algorithm~\ref{alg:SATE_ddp_general}} The following theorem establishes privacy guarantees for the framework:
\begin{theorem}\label{thm:SATE_ddp}
Let $\mcal{M}_1$ and $\mcal{M}_2$ be local randomizers for the first and second moments of $X_i$. Assume for all $n^* \in [n]$, $\mcal{M}_j(\cdot, n^*)$ satisfies $(\alpha, \varepsilon_j(\alpha))$-distributed R\'enyi DP for $j \in \{1, 2\}$. Then, Algorithm~\ref{alg:SATE_ddp_general} is $(\alpha, \varepsilon_1(\alpha)+\varepsilon_2(\alpha))$-R\'enyi DP.
\end{theorem}
\textbf{Proof.}
Since both $\hat{\Delta}_\msf{s}$ and $\hat{\sigma}_\msf{s}$ are functions of $\hat{\mu}_c, \hat{\mu}_t, \hat{s}^2_c,$ and $\hat{s}^2_t$, we only need to ensure their R\'enyi DP due to the post-processing properties of DP. The R\'enyi DP follows from a simple application of the composition theorem for R\'enyi DP \cite{mironov2017renyi}. \qedwhite

Note that although $\mcal{M}_1$ and $\mcal{M}_2$ are invoked twice in Algorithm~\ref{alg:SATE_ddp_general}, we only pay the privacy penalty once since one of the test or control groups remains the same for two neighboring datasets $\mb{y}$ and $\mb{y}'$. 

\subsection{Asymptotic Coverage Guarantees} 
Next, we claim that
Algorithm~\ref{alg:SATE_ddp_general} gives a $(1-\alpha)$-CI asymptotically.
\begin{assumption}\label{assumption:variance}
Assume the estimator $\hat{\mu}_{j}$, $j\in \{c, t\}$, are of an additive structure. 
That is, 
$\hat{\mu}_t = \sum_{i\in S_t} \hat{\mu}(\mcal{M}_1(X_i))$ and $\hat{\mu}_c = \sum_{i\in S_c} \hat{\mu}(\mcal{M}_1(X_i))$, where $\hat{\mu}\lp\mcal{M}_1(x_i, n^*)\rp$ gives an unbiased estimator, independent of $T_i$, on $x_i$ with variance bounded by $\sigma^2_1(n^*, \varepsilon)$\footnote{Indeed, we can relax the unbiasedness assumption and only require $\E\lb \hat{\mu}\lp \mcal{M}_1(x_i, n^*) \rp\rb = o(\frac{1}{n})$.};
\end{assumption}
\begin{assumption}
Assume $\hat{s}^2_c$ and $\hat{s}^2_t$ defined in Algorithm~\ref{alg:SATE_ddp_general} yield consistent estimation on the sample variances $s^2_c \eqDef \frac{1}{n-1}\sum_{i\in [n]} \lp y_i(c) - \bar{y}(c) \rp$ and $s^2_t \eqDef \frac{1}{n-1}\sum_{i\in [n]} \lp y_i(t) - \bar{y}(t) \rp$, respectively. That is, 
$ \hat{s}^2_c\lp \sum_{i \in S_c} \mcal{M}_2(X_i^2)\rp \pra \hat{\mu}_c$ as $n \ra \infty$ (and so does $\hat{s}_t$).
\label{assumption:consistent}
\end{assumption}
\begin{theorem}\label{thm:power_of_alg}
     Let the calibration term (which depends on $\mcal{M}_1$) be 
    \begin{equation}\label{eq:calibration}
        \sigma^2_\msf{pr}(n_c, n_t, \varepsilon) \eqDef \frac{n}{n_c}\sigma^2_1(n_c, \varepsilon) + \frac{n}{n_t}\sigma^2_1(n_t, \varepsilon).
    \end{equation}
    Then, under assumptions~\ref{assumption:variance} and \ref{assumption:consistent}, Algorithm~\ref{alg:SATE_ddp_general} gives a $(1-\alpha)$-confidence interval of SATE or PATE.
\end{theorem}

The proof of Theorem~\ref{thm:power_of_alg} can be found in Appendix~\ref{sec:additional_proofs}. We make a few remarks. First, in Algorithm~\ref{alg:SATE_ddp_general}, the CIs of SATE and PATE take slightly different forms. This is because the variance of SATE $\sigma^2_s$ depends on the sample covariance $s_{tc}$, which is an unidentifiable quantity. Thus, we can obtain a conservative upper bound $\hat{\sigma}_\msf{s}$. On the other hand, when the objective is to estimate PATE, the variance of the estimator does not depend on the covariance term, and thus $\hat{\sigma}^2_\msf{p}$ yields an unbiased estimator on the variance. 

Second, in order to determine a suitable treatment assignment size, denoted as $n_c$ and $n_t$, we can observe that the average length of confidence intervals (CIs) is influenced by two main factors:
\begin{align*}
    &\frac{\hat{\sigma}_s + \sigma^2_\msf{pr}(n_c, n_t, \varepsilon)}{n}
    \approx \underbrace{\frac{n_cn_t}{n}\lp \frac{\hat{s}_c}{n_c} + \frac{\hat{s}_t}{n_t} \rp^2}_{\text{(a)}} + \underbrace{\frac{\sigma^2_1(n_c, \varepsilon)}{n_c} + \frac{\sigma^2_1(n_t, \varepsilon)}{n_t}}_{\text{(b)}}.
\end{align*}
The first term (a) depends on the sample variances. 
To minimize this term, we should set $n_c$ and $n_t$ proportional to the sample variances of the control and treatment groups. 
However, since the sample variances are often unknown, estimating them requires additional samples and a privacy budget. 
On the other hand, the second term (b) represents the impact of DP noise. 
It is important to note that for a given value of $\varepsilon$, the variance of DP noise typically scales as $O\left(\frac{1}{n\min(\varepsilon, \varepsilon^2)}\right)$ (as we will see in the next section). 
Therefore, if either $n_c$ or $n_t$ is set too small, this term may dominate the total variance.

Apart from determining $n_c$ and $n_t$, another crucial question is how to allocate the privacy budget for estimating the first and second moments (i.e., the privacy used in $\mcal{M}_1$ and $\mcal{M}_2$). Allocating a significant portion of the privacy budget to estimate the mean (or difference-in-means) estimator can result in a relatively confident estimate of the Average Treatment Effect (ATE). 
However, this allocation may lead to inaccuracies in estimating the variance, affecting the accuracy of plug-in estimators for constructing CIs. 
In such cases, a more conservative estimate may be required to compute the CIs.
To address this issue, in the next section, we introduce non-asymptotic bounds that yield provable and more conservative coverage guarantees for specific mechanisms.


\section{Discrete DP Mechanisms for SecAgg}\label{sec:discrete_mechanisms}
In this section, we introduce discrete mechanisms that can be combined with secure aggregation for causal inference, which fall into two classes.

\paragraph{1. Additive Noise Mechanisms:} These mechanisms involve the addition of discrete noise approximating continuous Gaussian noise. In this approach, each local observable sample $X_i$ is first quantized into a discrete domain and then perturbed by adding appropriate discrete random noise. 
Candidate noise distributions include Binomial  \citep{agarwal2018cpsgd}, discrete Gaussian  \citep{canonne2020discrete, kairouz2021distributed}, and Skellam \citep{agarwal2021skellam}.

\paragraph{2. Randomized Response Mechanisms:} This class of mechanisms is based on the concept of randomized response introduced by \cite{warner1965randomized}. 
In these mechanisms, each sample $X_i$ is locally quantized into a binary value, and randomized response is applied multiple times with an appropriate cross-over probability determined by $\varepsilon$. The results of the randomized responses are summed together. Equivalently, this scheme can be viewed as having each client encode its message as a parameter of a Binomial random variable, sending a sample of it to the server. 
The decoded output follows a Poisson-binomial distribution, resulting in the Poisson-binomial mechanism (PBM). 
Note that since the output space of PBM is finite, it is compatible with secure aggregation, and hence no modular-clipping is required. Therefore, the resulting estimator is unbiased, while all of the additive noise mechanisms inevitably have to introduce small biases. 

For brevity, we only present the results of randomized response mechanisms here, and the analysis of additive noise mechanisms is similar.

\subsection{Difference-in-mean estimator with the Poisson-binomial mechanism}
\begin{algorithm}[H]
   \caption{The Poisson Binomial Mechanism}
   \label{alg:scalar_pbm}
\begin{algorithmic}
    \State {\bfseries Input:} $c > 0$, $x_i \in [-R, R]$
    \State {\bfseries Parameters:} $\theta \in [0, \frac{1}{4}]$, $m \in \mbb{N}$
    \State Re-scaling $x_i$:
    $ p_i \eqDef \frac{\theta}{R}x_i+\frac{1}{2}$.
    \State Privatization:
    $ Z_i \eqDef \msf{Binom}\lp m, p_i \rp \in \mbb{Z}_m.$
    \State {\bfseries Return:} $Z_i$
\end{algorithmic}
\end{algorithm}
Next, we describe and analyze our distributed DP scheme based on the Poisson-binomial mechanism (PBM) \citep{chen2022poisson}. 
We make the same assumption that the potential outcome space $\mcal{Y}$ is a bounded interval and is known ahead of time. Without loss of generality, we let $\mcal{Y} = [-R, R]$ for some $R > 0$\footnote{Here we assume $R>0$ is known beforehand, which is often the case. 
When $R$ is unknown, we may need to estimate it through private range/quantile queries.}. 
Per Theorem~\ref{thm:SATE_ddp}, our goal here is to specify the R\'enyi DP guarantees and the variance of the scheme.

The local randomizer $\mcal{M}_{\msf{PBM}}$ is described in Algorithm~\ref{alg:scalar_pbm}, which consists of two main steps:  1) first mapping $x_i$ into $\lb \frac{1}{2}-\theta, \frac{1}{2}+\theta \rb$ by  $p_i \eqDef \frac{1}{2}+\frac{\theta}{R}x_i$, and then 2) generating a binomial random variable $Z_i \sim \msf{Binom}(m, p_i)$.

Upon securely aggregating $\sum_i Z_i$, the server can obtain an unbiased estimator on $\mu = \sum_i x_i$ as 
\begin{equation}\label{eq:pbm_estimator}
\textstyle
    \hat{\mu}\lp \sum_i Z_i \rp \eqDef \frac{R}{nm\theta}\lp \sum_i Z_i- \frac{m}{2}\rp
\end{equation}
(recall that the server can only learn $\sum_i Z_i$ but not individual $Z_i$'s). In the following theorem, we summarize the privacy and the variance of PBM for a given set of parameters $(m, \theta)$.
\begin{theorem}[\cite{chen2022poisson}]\label{thm:pbm}
    Let $\hat{\mu}$ be the estimator from \eqref{eq:pbm_estimator}. Via Assumption~\ref{assumption:bdd_x}, for any $\theta\!\in\![0,\frac14]$
    \begin{itemize}
        \item $\hat{\mu}$ yields an \emph{unbiased} estimate on $\mu$ with variance at most $\frac{R^2}{4nm\theta^2}$.
        \item Algorithm~\ref{alg:scalar_pbm}, together with SecAgg \citep{bonawitz2016practical}, satisfies $(\alpha, \varepsilon(\alpha))$-R\'enyi DP for any $\alpha > 1$ and
    \begin{equation}\textstyle
        \varepsilon(\alpha) \geq C \lp \frac{\theta^2}{(1-2\theta)^4}\rp\frac{\alpha m}{n},
    \end{equation}
    where $C > 0$ is a universal constant.
    \end{itemize}
\end{theorem}
From this, we can re-write the MSE (i.e., the variance) as $\Var\lp \hat{\mu} \rp \leq \frac{R^2}{4nm\theta^2} = O\lp \frac{R^2\alpha}{n^2\varepsilon(\alpha)} \rp$.

Since $Z_i \leq m$ and thus $\sum_i Z_i \leq nm$, we set the modulo space $M = nm+1$ to avoid overflow (recall that $M$ is the size of the finite group SecAgg operates on). Therefore, the communication cost of Algorithm~\ref{alg:scalar_pbm} is $\log M \approx \log n + \log m$ bits per client. In addition, unlike in the additive mechanisms where the noise support is typically unbounded, there is no need to apply modular clipping, and thus $\hat{\mu}$ is unbiased.

    A limitation of the PBM approach is that 
    the mechanism was designed for federated learning tasks where local messages are high-dimensional vectors (i.e., model updates) and the number of per-round users is small (usually less than $10^3$)~\citep{chen2022poisson}. 
    However, in the design of the experiments, the number of tests can easily exceed millions, and the privacy accounting algorithm in \citet{chen2022poisson} becomes infeasible. 
    In this work, we develop new efficiently computable bounds on the R\'enyi DP of PBM that are within  1\% greater of the actual privacy loss, described in  Appendix~\ref{app:pbm_accounting}.

Next, we construct the mechanisms $\mcal{M}_1(\cdot, n^*)$ and $\mcal{M}_2(\cdot, n^*)$ used in Algorithm~\ref{alg:SATE_ddp_general}. Let $(m_{1, c}, \theta_{1,c})$, $(m_{1, t}, \theta_{1, t})$ be the parameters of PBM used for estimating the mean of the control and test groups respectively. Similarly, let $(m_{2, c}, \theta_{2,c})$, $(m_{2, t}, \theta_{2, t})$ be the parameters used in estimating the second moments of the two groups. Then according to Theorem~\ref{thm:pbm}, the privacy losses of $\mcal{M}_1(\cdot, n_c)$ and $\mcal{M}_1(\cdot, n_t)$ are $O\lp \frac{\alpha\theta_{1, c}^2m_{1, c}}{n_c} \rp$ and $O\lp \frac{\alpha\theta_{1, t}^2m_{1, t}}{n_t} \rp$\footnote{Note that although here we present an asymptotic form of the privacy losses, in our experiments we can numerically compute the accurate privacy budgets.}, and the privacy losses of $\mcal{M}_2(\cdot, n_c)$ and $\mcal{M}_2(\cdot, n_t)$ are 
$O\lp \frac{\alpha\theta_{2, c}^2m_{2, c}}{n_c} \rp$ and $O\lp \frac{\alpha\theta_{2, t}^2m_{2, t}}{n_t} \rp$.
Therefore, combining Theorem~\ref{thm:pbm} with Theorem~\ref{thm:SATE_ddp}, we summarize the guarantees of PBM in the following corollary:
\begin{corollary}
    Let $\mcal{M}_1$ and $\mcal{M}_2$ be implemented with PBM with parameters $(m_{1, c}, \theta_{1,c})$, $(m_{1, t}, \theta_{1, t})$, $(m_{2, c}, \theta_{2,c})$, and $(m_{2, t}, \theta_{2, t})$ respecitvely. Then 
    \begin{enumerate}
        \item Alg.~\ref{alg:SATE_ddp_general} is $(\alpha, \varepsilon(\alpha))$-R\'enyi DP for all $\alpha > 1$ and
        \begin{align*} \hspace*{-6mm}\textstyle
            \varepsilon(\alpha) = O\Bigg( \alpha\Big( \frac{\theta_{1, c}^2m_{1, c}}{n_c} \textstyle+ \frac{\theta_{1, t}^2m_{1, t}}{n_t}+ \frac{\theta_{2, c}^2m_{2, c}}{n_c}+ \frac{\theta_{2, c}^2m_{2, c}}{n_c}  \Big) \Bigg).
        \end{align*}
        \item  The average width of the $(1-\alpha)$-CI is 
        $$ \textstyle O\lp z_{1-\frac{\alpha}{2}}\cdot \sqrt{ \frac{s_c^2}{n_c}+\frac{s_t^2}{n_t}+\frac{c^2}{n_tm_{1,t}\theta_{1,c}^2}+\frac{c^2}{n_tm_{1,t}\theta_{1,t}^2}} \rp$$ for SATE, and for PATE it is
        {\small
        \begin{align*} \hspace*{-6mm}
            O\Bigg(z_{1-\frac{\alpha}{2}}\cdot 
            \textstyle \sqrt{\frac{\Var\lp P_0 \rp}{n_c}+\frac{\Var\lp P_1 \rp}{n_t}+\frac{c^2}{n_tm_{1,t}\theta_{1,c}^2}+\frac{c^2}{n_tm_{1,t}\theta_{1,t}^2}} \Bigg).
        \end{align*}
        }
    \end{enumerate}
    \end{corollary}

Note that in the above expression, the parameters of $\mcal{M}_2$ do not impact the (asymptotic) width of the confidence intervals (CIs). This is because as long as we can derive a consistent estimator for the sample variances, we can compute CIs accordingly. Therefore, one should allocate the maximum possible privacy budget to $\mcal{M}_1$. In practice (Section~\ref{sec:experiments}), we set the privacy budget for $\mcal{M}_1$ to be $0.99$ of the total privacy allocation.

\paragraph{Parameter selection.}
In order to satisfy a $(\varepsilon, \delta)$-DP, guarantee, we select $$ \textstyle\frac{\theta_{1, c}^2m_{1, c}}{n_c} \approx \frac{\theta_{1, t}^2m_{1, t}}{n_t} = O_\delta\lp \varepsilon^2 \rp,$$
which means that the average width of the CIs is
$O\!\lp\!z_{1-\frac{\alpha}{2}}\!\cdot\!\sqrt{ \frac{s_c^2}{n_c}+\frac{s_t^2}{n_t}+\frac{c^2}{\varepsilon^2}\lp \frac{1}{n_t^2}+\frac{1}{n_c^2}\rp}\rp$ for SATE, or $O\!\lp\!z_{1-\frac{\alpha}{2}}\!\cdot\!\sqrt{ \frac{\Var\lp P_0 \rp}{n_c}\!+\!\frac{\Var\lp P_1 \rp}{n_t}\!+\!\frac{c^2}{\varepsilon^2}\lp \frac{1}{n_t^2}\!+\!\frac{1}{n_c^2}\rp} \rp$ for PATE.

\begin{table*}[th]
\centering
\caption{Average widths and coverages of $90\%$-confidence intervals for PATE.}
\label{tab:exp1}
\label{tab:exp_PATE}
\scriptsize
\setlength{\tabcolsep}{4.0pt}
\begin{tabular}{@{}cccccccccc@{}}
\toprule
\multicolumn{1}{c}{} &&
  $\varepsilon = 0.1$ &
  $\varepsilon = 0.4$ &
  $\varepsilon = 0.7$ &
  $\varepsilon = 1.0$ &
  $\varepsilon = 1.3$ &
  $\varepsilon = 1.6$ &
  $\varepsilon = 1.9$ &
  $\varepsilon = \infty$ \\ \midrule
\multirow{2}{*}{None private} &
  Coverage (90\% CI) &
  - &-&-&-&-&- &- &0.902 \\
 &
  Width (90\% CI) &
  - &-&-&-&-&- &- &2.08$\cdot10^{-3}$ \\\midrule

\multirow{2}{*}{Central Gaussian} &
  Coverage (90\% CI) &
  0.899 & 0.901 & 0.902 & 0.899 & 0.897 & 0.897 & 0.899 &- \\
 &
  Width (90\% CI) &
  0.771 & 0.189 & 0.110 &  0.078  & 0.063 & 0.053 & 0.044 &-\\\midrule

  \multirow{2}{*}{PBM ($m=256$)} &
  Coverage (90\% CI) &
  0.898 & 0.897  & 0.903 & 0.900 & 0.902 & 0.899 & 0.903 &- \\
 &
  Width (90\% CI) &
  0.772 &
  0.200 &
  0.119 &
  0.085 &
  0.067 &
  0.056 &
  0.048 &- \\\midrule
  
  \multirow{2}{*}{PBM ($m=1024$)} &
  Coverage (90\% CI) &
  0.903 & 0.899 & 0.899  & 0.899 & 0.902 & 0.898 & 0.900 &-\\
 &
 Width (90\% CI) &
  0.772 & 0.199 & 0.118 & 0.085 & 0.066 & 0.055 & 0.047 &-\\ 

  \bottomrule
\end{tabular}
\end{table*}

\subsection{Non-asymptotic coverage guarantees}
In addition to the asymptotic CIs based on the central limit theorem, which are accurate only when $n_c$ and $n_t$ are large, we provide non-asymptotic CIs for estimating SATE and PATE based on variants of empirical Berstein inequalities. 
For ease of presentation, in the rest of this section, we assume $n_c = n_t = n/2$, but all of the results can be easily adapted to general cases. 
We first present the non-asymptotic bound for PATE.
\begin{theorem}[Simplified]\label{thm:pate_CIs}
    Let $\mcal{M}_1$ and $\mcal{M}_2$ be PBM (Algorithm~\ref{alg:scalar_pbm}) with parameter $(m_1, \theta_1)$ and $(m_2, \theta_2)$. Let $\hat{\sigma}^2_\msf{p}$ be defined as in Algorithm~\ref{alg:SATE_ddp_general}. Then under Assumption~\ref{assumption:bdd_x}, it holds that
    \begin{align*}
        \Pr\lbp \Delta_\msf{p} \in  \hat{\Delta}_\msf{p} \pm \lp \sqrt{2\hat{\sigma}^2_\msf{p} \log(2.01/\delta)} + \gamma\rp \rbp \geq 1-\delta,
    \end{align*}
    where $\gamma = O(1/n)$ when $\varepsilon_1(\alpha) = C_1 \frac{\alpha m_1 \theta_1^2}{n}$ and $\varepsilon_2(\alpha) = C_2 \frac{\alpha m_2 \theta_2^2}{n}$ are constants\footnote{We provide the higher-order terms and constants of $\gamma$ in Theorem~\ref{thm:pate_CIs_detailed} in Appendix~\ref{appendix:proof_of_pate_berstein}}.
\end{theorem}
The above theorem is proved via the empirical Berstein inequality~\citep{maurer2009empirical} along with incorporating the tail bounds of the Poisson binomial mechanism. 
The same analysis can be applied to other additive mechanisms (such as the Skellam or discrete Gaussian noise), though these mechanisms may not yield an unbiased estimator. 
The detailed proof can be found in Appendix~\ref{appendix:proof_of_pate_berstein}.

For a given privacy budget $\varepsilon(\alpha)$, Theorem~\ref{thm:pate_CIs} suggests a way to allocate privacy budgets (determined by $(m_1, \theta_1)$ and $(m_2, \theta_2)$) to minimize the width of CIs (i.e., $2\tau$). 
Specifically, if we split the total privacy budget $\varepsilon(\alpha)$ into $\varepsilon_1(\alpha)$ and $\varepsilon_2(\alpha)$ such that $\mcal{M}_1$ and $\mcal{M}_2$ satisfy $\varepsilon_1(\alpha)$ and $\varepsilon_2(\alpha)$ R\'enyi DP respectively, then we have the following proposition:
\begin{proposition}\label{prop:pate_split}
    Make the same assumptions as Theorem~\ref{thm:pate_CIs}. Assume $\mcal{M}_1$ and $\mcal{M}_2$ satisfies $(\alpha, \varepsilon_1(\alpha))$ and $(\alpha, \varepsilon_2(\alpha))$ R\'enyi DP. Let the sample variance $\hat{s}^2_c$ and $\hat{s}^2_t$ be constant and do not scale with $n$. Then the non-asymptotic CIs in Theorem~\ref{thm:pate_CIs} has width
    {\small
    \begin{align*} \textstyle
        \Theta_\delta\bigg( &\textstyle\sqrt{\frac{\hat{s}^2_t+\hat{s}^2_c}{n}} + \frac{R}{n}\lp 1+\sqrt{\frac{\alpha}{\varepsilon_1(\alpha)}} \rp + \frac{R^2\sqrt{\alpha/\varepsilon_2(\alpha)} + R\sqrt{\alpha/\varepsilon_1(\alpha)}}{n^{1.5}\sqrt{\hat{s}^2_t + \hat{s}^2_c}} \bigg).
    \end{align*}
    }
\end{proposition}

We provide some insights regarding Proposition~\ref{prop:pate_split}. First, it is important to note that the DP noise only impacts the smaller terms (i.e., $O(1/n)$). The first-order term $\sqrt{(\hat{s}^2_t + \hat{s}^2_c)/n}$ remains consistent with the asymptotic confidence intervals. Additionally, when considering the allocation of privacy budget $\varepsilon_2(\alpha)$  for estimating sample variance, it exerts a comparatively lesser influence on the confidence intervals in contrast to $\varepsilon_1(\alpha)$ since $\varepsilon_2(\alpha)$ only plays a role in the $O(1/n^{1.5})$ term. This observation supports our intuition that allocating more privacy budget to $\mcal{M}_1$ is advisable when dealing with sufficiently large values of n.

It is also essential to emphasize that we do not advocate the use of non-asymptotic confidence bounds (as presented in Theorem~\ref{thm:pate_CIs}) over the asymptotic ones (Theorem~\ref{thm:power_of_alg}). This is because non-asymptotic bounds may still be overly conservative. Instead, Theorem~\ref{thm:pate_CIs} should be utilized as a guideline for allocating privacy budgets when dealing with finite sample sizes.

Finally, the same non-asymptotic CIs hold for SATE.
\begin{theorem}\label{thm:sate_CIs}
    Theorem~\ref{thm:pate_CIs} holds for SATE by replacing $\hat{\sigma}^2_\msf{p}$, $\Delta_\msf{p}$, and $\hat{\Delta}_\msf{p}$ with $\hat{\sigma}^2_\msf{s}$, $\Delta_\msf{s}$, and $\hat{\Delta}_\msf{s}$.
\end{theorem}
The proof is more involved as it requires a sample-without-replacement version of Berstein inequality. We leave the details in Appendix~\ref{appendix:proof_sate_berstein}.

\section{Experiments}\label{sec:experiments}

In this section, we provide empirical evaluations for our proposed framework.

\paragraph{Experiment Setup.} We generate the potential outcomes according to truncated Gaussian distributions. Specifically, we set the (population) ATE to be $0.2$ and generate $Y_i(c) \diid N(-0.1, \sigma_p^2)$ and $Y_i(t) \diid N(0.1, \sigma_p^2)$, with $\sigma_p = 0.05$. We truncate both $Y_i(c)$ and $Y_i(t)$ to $[-1, 1]$. We divide the sample size $n = 10^4$ equally into test and control groups (i.e., $n_c = n_t = 5\cdot 10^3$).
We set the confidence level to be $90\%$, simulate for $N = 10000$ rounds, and compute the empirical coverage ratio, i.e., the number of times that the true PATE lies in the estimated CIs. 

\paragraph{Baselines.} We compare the proposed distributed DP method, based on PBM (labeled as ``PBM'') with (1) the non-private difference-in-mean CIs and (2) the Central DP baseline (where we collect all observable samples and add Gaussian noise to the difference-in-mean estimator). For PBM, we compare different output sizes $m$ (recall that $m$ determines the per-user communication cost). We report the average widths of the $90\%$-CIs, as well as the empirical coverage rates.

From Table~\ref{tab:exp1}, we see that the widths of CIs are largely determined by the DP noise and the corresponding privacy levels. However, the CI widths of PBM are very close to the Central Gaussian mechanism, indicating that the price of adopting secure aggregation is small. Due to space limitations, we provide more detailed experimental results in the appendix, including the CIs for SATE and under different data distributions.

\newpage
\bibliography{references}
\newpage

\onecolumn
\newpage
\appendix
\section{Practical privacy accounting for PBM}\label{app:pbm_accounting}
In this section, we improve the efficiency of the privacy accounting mechanism \cite{chen2022poisson}, which are originally designed for small sample and finite field sizes (usually when $n, m \leq 10^3$) due to the batch-SGD and the natural computation and communication constraints of using secure aggregation.

Following from the proof of Theorem~3.3 in \cite{chen2022poisson}, for any set of parameters $(m, n, \theta, \alpha)$, $\varepsilon(\alpha)$ can be expressed as
\begin{align}\label{eq:extreme_1}
&\max_{t_1, t_2 \in [m\cdot n], \, |t_1 - t_2| \leq m} D_\alpha\Big( P_{\msf{Binom}\lp t_1, \frac{1}{2}-\theta\rp + \msf{Binom}\lp mn-t_1, \frac{1}{2}+\theta\rp}\big\Vert P_{\msf{Binom}\lp t_2, \frac{1}{2}-\theta\rp + \msf{Binom}\lp mn-t_2, \frac{1}{2}+\theta\rp} \Big).
\end{align}

In \cite{chen2022poisson}, it is shown that the maximum of \eqref{eq:extreme_1} occurs at $(t_1, t_2)=(0, m)$, which suggests the following (exact) privacy accounting mechanism in Algorithm~\ref{alg:accounting_exact}. 

\begin{algorithm}[H]
   \caption{Exact privacy accounting.}
   \label{alg:accounting_exact}
\begin{algorithmic}
    \State {\bfseries Input:} $n, m, \theta, \alpha$
    \State {\bfseries Return:} $\varepsilon(\alpha)$
    \State
    $ P_1 \la \msf{Binom}(mn, \frac{1}{2}-\theta)$ \algorithmiccomment{$P_1$ is a $mn+1$-dim vector.}
    \State $P_2 \la \msf{Binom}(m(n-1), \frac{1}{2}-\theta)$ 
    \State $P'_2 \la \msf{Binom}(m, \frac{1}{2}+\theta)$ 
    \State $P_2 \la P_2 * P'_2 $ \algorithmiccomment{$*$ denotes the convolution operator.}
    \State $ \varepsilon(\alpha) \la \frac{1}{\alpha-1}\log\lp \msf{sum}\lp \frac{P_1^\alpha}{P_2^{\alpha-1}} \rp \rp$ \algorithmiccomment{$\msf{sum}$ and $(\cdot)^\alpha$ are performed coordinate-wisely.}
\end{algorithmic}
\end{algorithm}

Note that the accounting involves binomial coefficients with large $n$, so in practice, all computations should be done in the log space to ensure computation stability, as described in Algorithm~\ref{alg:accounting_exact_log}. The computation bottlenecks of Algorithm~\ref{alg:accounting_exact} and Algorithm~\ref{alg:accounting_exact_log} are at the convolution operation, which, when computed via fast Fourier transform, takes $\tilde{O}(mn)$ time.

\begin{algorithm}[H]
   \caption{Exact privacy accounting over the log space.}
   \label{alg:accounting_exact_log}
\begin{algorithmic}
    \State {\bfseries Input:} $n, m, \theta, \alpha$
    \State {\bfseries Return:} $\varepsilon(\alpha)$
    \State
    $ \msf{logP_1} \la \log\lp \msf{Binom}(mn, \frac{1}{2}-\theta)\rp$ 
    \State $\msf{logP_2} \la \log \lp \msf{Binom}(m(n-1), \frac{1}{2}-\theta) \rp$ 
    \State $\msf{logP'_2}\la \msf{Binom}(m, \frac{1}{2}+\theta)$ 
    \State $\msf{logP_2} \la \msf{logP_2 } \,\tilde{*} \,\msf{ logP'_2} $ \algorithmiccomment{$\tilde{*}$ denotes the convolution operator \emph{over the log space}.}
    \State $ \varepsilon(\alpha) \la \frac{1}{\alpha-1} \msf{logexpsum}\lp \alpha\cdot\msf{logP_1} +(1-\alpha)\cdot\msf{logP_2} \rp$ 
\end{algorithmic}
\end{algorithm}

\subsection{Approximation for large $n$ and $m$}
Unfortunately, in most private analytic or causal inference tasks, the number of samples $n$ can be up to millions (and $m$ may be up to thousands), making the $\tilde{O}(mn)$ time complexity of the above algorithms infeasible. To address this issue, we propose to account for the privacy loss via the following upper bound based on a data processing inequality:
\begin{align}\label{eq:extreme_2}
\eqref{eq:extreme_1} \leq \max_{k \in [n-1]} m\cdot D_\alpha\Big( P_{\msf{Binom}\lp 1+k, \frac{1}{2}-\theta\rp + \msf{Binom}\lp n-k-1, \frac{1}{2}+\theta\rp}\big\Vert P_{\msf{Binom}\lp k, \frac{1}{2}-\theta\rp + \msf{Binom}\lp n-k, \frac{1}{2}+\theta\rp} \Big).
\end{align}
Although \eqref{eq:extreme_2} is always strictly greater than the exact privacy loss \eqref{eq:extreme_1}, when either $m$ or $n$ is large, the approximation error in $\varepsilon(\alpha)$ is negligible. For instance, when $n=100$ and $\alpha=2$, the approximation error is less than $0.1\%$. By leveraging \eqref{eq:extreme_2}, we arrive at the following approximate privacy accounting algorithm, which reduces the computational complexity to $O(n)$:

\begin{algorithm}[H]
   \caption{Efficient approximate privacy.}
   \label{alg:accounting_approx_log}
\begin{algorithmic}
    \State {\bfseries Input:} $n, m, \theta, \alpha$
    \State
    $ \msf{logP_1} \la \log\lp \msf{Binom}(n, \frac{1}{2}-\theta)\rp$ 
    \State $\msf{logP_2} \la \log \lp \msf{Binom}(n-1, \frac{1}{2}-\theta) \rp$ 
    \State $\msf{logP'_2}\la \msf{Ber}(\frac{1}{2}+\theta)$ 
    \State $\msf{logP_2} \la \msf{logP_2 } \,\tilde{*} \,\msf{ logP'_2} $ \algorithmiccomment{$\tilde{*}$ denotes the convolution operator \emph{over the log space}.}
    \State $ \varepsilon(\alpha) \la \frac{1}{\alpha-1} \msf{logexpsum}\lp \alpha\cdot\msf{logP_1} +(1-\alpha)\cdot\msf{logP_2} \rp$.

    \State {\bfseries Return:} $m\varepsilon(\alpha)$
\end{algorithmic}
\end{algorithm}

In our experiments, we account the R\'enyi DP according to Algorithm~\ref{alg:accounting_approx_log} and convert the $(\alpha, \varepsilon(\alpha))$-R\'enyi DP to $(\varepsilon, \delta)$-DP via the conversion lemma given by \cite{canonne2020discrete}.

\section{Additional experiments}
In this section, we provide more complete experimental results to demonstrate the utility of our proposed framework.

\subsection{Gaussian potential outcomes}\label{subsec:exp_gaussian}

In the first set of examples, we consider random treatment effects, where the potential outcomes before and after the treatment are normally distributed: $Y_i(0)\diid N(\mu_0, \sigma)$ and $Y_i(1)\diid N(\mu_1, \sigma)$. Under this distributional assumption, the PATE is defined as $\Delta_\msf{p} \eqDef \mu_1 - \mu_0$, while the SATE is $\Delta_\msf{s} \eqDef \frac{1}{n_t}\sum_i Y_i(1) - \frac{1}{n_t}\sum_i Y_i(0)$, where $n_c$ and $n_t$ represent the numbers of the control and test groups.

In the experiments, we set $n_c = n_t = 10^3$, $\Delta_\msf{p} = 0.2$, and the noise level $\sigma = 0.01$. For each set of parameters of the privatization mechanisms, we set the confidence level to be $90\%$, simulate for $N = 10000$ rounds, and report the average widths of CIs and the empirical coverage ratios (i.e., the number of times that the true PATE lies within the estimated CIs).

In Table~\ref{tbl:exp_SATE_n_1000_gauss}, we observe that without privacy constraints, we obtain tight CIs with a significantly higher coverage ratio than required. Specifically, we achieve a coverage ratio of 0.98 compared to the requested 0.9 coverage ratio under a 90\% confidence constraint\footnote{Note that when estimating the confidence intervals of the difference-in-mean estimator for SATE, the true variance is unidentifiable. Therefore, we can only use an upper bound to obtain a conservative interval, as discussed in the proof of Theorem~\ref{thm:SATE_ddp}.}. The issue of being overly conservative, however, vanishes under DP, since the DP noise dominates the total uncertainty and is much larger than the sampling variance.

Comparing the non-private setting, we found that the width of the private CIs is significantly larger than the non-private one, indicating that the DP noise is much larger than the sampling noise. Unfortunately, this is the price we need to pay. However, the CI widths of the centralized Gaussian mechanism are roughly the same as the width of PBM. The difference to the Gaussian mechanism is negligible when $n$ and $m$ are large enough. In Table~\ref{tbl:exp_SATE_n_1000_gauss}, we can see that when $n = 1000$, setting $m=256$ is sufficient to achieve the same performance as the centralized Gaussian mechanism. This implies that although the price for achieving DP is indispensable, the price for adopting secure aggregation to remove the trust toward the server can be made arbitrary small, as long as we are willing to slightly increase the communication costs (which are dictated by the finite field size $m$).

{\small
\begin{table}
\centering
\caption{Average width and coverage of $90\%$-confidence intervals for SATE. Gaussian potential outcomes with $n=10^3$.}
\vspace{0.1 cm}
\label{tbl:exp_SATE_n_1000_gauss}
\tiny
\setlength{\tabcolsep}{4.0pt}
\begin{tabular}{@{}cccccccc@{}}
\toprule

\multirow{2}{*}{Non-private} 

& \multicolumn{7}{c}{0.980}\\
& \multicolumn{7}{c}{ 0.002 $\pm$  3.25e-05}
\\ \midrule
\multicolumn{1}{c}{} &
  $\varepsilon = 0.1$ &
  $\varepsilon = 0.4$ &
  $\varepsilon = 0.7$ &
  $\varepsilon = 1.0$ &
  $\varepsilon = 1.3$ &
  $\varepsilon = 1.6$ &
  $\varepsilon = 1.9$ \\ \midrule
\multirow{2}{*}{Central Gaussian} 
&  0.899
&  0.897
&  0.899
&  0.900
&  0.901
&  0.897
&  0.899
\\
&  0.771 $\pm$  1.26e-07
&  0.199 $\pm$  4.87e-07
&  0.118 $\pm$  8.40e-07
&  0.084 $\pm$  1.15e-06
&  0.066 $\pm$  1.45e-06
&  0.055 $\pm$  1.78e-06
&  0.047 $\pm$  2.08e-06 \\\midrule
  \multirow{2}{*}{PBM (m=256)} 
&  0.899
&  0.903
&  0.904
&  0.905
&  0.898
&  0.896
&  0.896  
  \\
& 0.772 $\pm$  1.26e-07
&  0.200 $\pm$  4.85e-07
&  0.119 $\pm$  8.34e-07
&  0.085 $\pm$  1.13e-06
&  0.067 $\pm$  1.42e-06
&  0.056 $\pm$  1.73e-06
&  0.048 $\pm$  2.00e-06
 \\\midrule
  
  \multirow{2}{*}{PBM (m=1024)} 
&  0.904
&  0.892
&  0.896
&  0.901
&  0.901
&  0.904
&  0.898
  \\
&  0.772 $\pm$  1.26e-07
&  0.199 $\pm$  4.83e-07
&  0.118 $\pm$  8.23e-07
&  0.085 $\pm$  1.15e-06
&  0.066 $\pm$  1.47e-06
&  0.055 $\pm$  1.76e-06
&  0.047 $\pm$  2.07e-06 
\\\midrule

  \multirow{2}{*}{PBM (m=2048)} 
&  0.896
&  0.902
&  0.899
&  0.903
&  0.897
&  0.904
&  0.896\\

& 0.772 $\pm$  1.27e-07
&  0.199 $\pm$  4.81e-07
&  0.118 $\pm$  8.16e-07
&  0.084 $\pm$  1.15e-06
&  0.066 $\pm$  1.45e-06
&  0.055 $\pm$  1.77e-06
&  0.047 $\pm$  2.08e-06\\
  \bottomrule
\end{tabular}
\end{table}
}

We can observe a similar trend when estimating the population level treatment effect (i.e., PATE). We see that when setting $m=256$, the width of CIs is almost the same as the the centralized Gaussian. A major difference compared to estimating SATE, however, is that the average converge ratio of the non-private setting becomes aligned with our target confidence level (i.e., 90\% in our setting). This is because the variance estimator of PATE given in Algorithm~\ref{alg:SATE_ddp_general} becomes unbiased since the unidentifiable term (i.e., the covariance) is cancelled out (see the proof given in Section~\ref{sec:additional_proofs} for more details).

{\small
\begin{table}[H]
\centering
\caption{Average width and coverage of $90\%$-confidence intervals for PATE. Gaussian potential outcomes with $n=10^3$.}
\vspace{0.1 cm}
\label{tab:exp_SATE}
\tiny
\setlength{\tabcolsep}{4.0pt}
\begin{tabular}{@{}cccccccc@{}}
\toprule

\multirow{2}{*}{Non-private} 

& \multicolumn{7}{c}{0.901}\\
& \multicolumn{7}{c}{0.002 $\pm$  3.24e-05}
\\ \midrule
\multicolumn{1}{c}{} &
  $\varepsilon = 0.1$ &
  $\varepsilon = 0.4$ &
  $\varepsilon = 0.7$ &
  $\varepsilon = 1.0$ &
  $\varepsilon = 1.3$ &
  $\varepsilon = 1.6$ &
  $\varepsilon = 1.9$ \\ \midrule
\multirow{2}{*}{Central Gaussian} 

&  0.905
&  0.895
&  0.899
&  0.902
&  0.904
&  0.899
&  0.899

\\
&  0.771 $\pm$  1.24e-07
&  0.199 $\pm$  4.85e-07
&  0.118 $\pm$  8.20e-07
&  0.084 $\pm$  1.16e-06
&  0.066 $\pm$  1.47e-06
&  0.055 $\pm$  1.78e-06
&  0.047 $\pm$  2.07e-06 \\\midrule
  \multirow{2}{*}{PBM (m=256)} 
&  0.902
&  0.900
&  0.900
&  0.903
&  0.906
&  0.900
&  0.903 
  \\
&  0.772 $\pm$  1.25e-07
&  0.200 $\pm$  4.84e-07
&  0.119 $\pm$  8.15e-07
&  0.085 $\pm$  1.15e-06
&  0.067 $\pm$  1.43e-06
&  0.056 $\pm$  1.72e-06
&  0.048 $\pm$  2.02e-06
 \\\midrule
  
  \multirow{2}{*}{PBM (m=1024)}

&  0.900
&  0.897
&  0.902
&  0.900
&  0.904
&  0.898
&  0.896

  \\
&  0.772 $\pm$  1.26e-07
&  0.199 $\pm$  4.85e-07
&  0.118 $\pm$  8.28e-07
&  0.085 $\pm$  1.17e-06
&  0.066 $\pm$  1.46e-06
&  0.055 $\pm$  1.77e-06
&  0.047 $\pm$  2.05e-06
\\\midrule

  \multirow{2}{*}{PBM (m=2048)} 

&  0.897
&  0.902
&  0.901
&  0.901
&  0.899
&  0.902
&  0.898
\\

&  0.772 $\pm$  1.24e-07
&  0.199 $\pm$  4.85e-07
&  0.118 $\pm$  8.19e-07
&  0.084 $\pm$  1.16e-06
&  0.066 $\pm$  1.47e-06
&  0.055 $\pm$  1.77e-06
&  0.047 $\pm$  2.06e-06\\
  \bottomrule
\end{tabular}
\end{table}
}

\subsection{Constant treatment effects}
In the second set of examples, we consider constant treatment effects. Specifically, we assume $Y_i(0)\diid \msf{uniform}(a, b)$ and $Y_i(1) = Y_i(0) + \Delta_\msf{s}$, where $\Delta_\msf{s}$ is a deterministic but unknown quantity that we want to estimate.

In the experiments, we set $n_c = n_t = 10^3$, $\Delta_\msf{s} = 0.2$, and $(a, b) = (-1, -0.8)$. For each set of parameters of the privatization mechanisms, we again set the confidence level to be $90\%$, simulate for $N = 10000$ rounds, and report the average widths of CIs and the empirical coverage ratios.

As shown in Table~\ref{tbl:exp_SATE_n_1000_constant} and Table~\ref{tbl:exp_PATE_n_1000_constant}, under the assumption of a constant ATE, estimating SATE and PATE is essentially the same, both theoretically and empirically. The coverage ratios for both PATE and SATE are accurate, in contrast to SATE with random ATE. Furthermore, we observe a similar trend as in the Gaussian outcomes, where PBM achieves a negligible error compared to the central Gaussian.

{\small
\begin{table}[H]
\centering
\caption{Average width and coverage of $90\%$-confidence intervals for SATE. Constant treatment effect with $n=10^3$.}
\vspace{0.1 cm}
\label{tbl:exp_SATE_n_1000_constant}
\tiny
\setlength{\tabcolsep}{4.0pt}
\begin{tabular}{@{}cccccccc@{}}
\toprule

\multirow{2}{*}{Non-private} 

& \multicolumn{7}{c}{0.897}\\
& \multicolumn{7}{c}{ 
  0.108 $\pm$  1.53e-03}
\\ \midrule
\multicolumn{1}{c}{} &
  $\varepsilon = 0.1$ &
  $\varepsilon = 0.4$ &
  $\varepsilon = 0.7$ &
  $\varepsilon = 1.0$ &
  $\varepsilon = 1.3$ &
  $\varepsilon = 1.6$ &
  $\varepsilon = 1.9$ \\ \midrule
\multirow{2}{*}{Central Gaussian} 

&  0.904
&  0.902
&  0.901
&  0.899
&  0.895
&  0.899
&  0.896

\\

&  0.779 $\pm$  2.11e-04
&  0.227 $\pm$  7.30e-04
&  0.160 $\pm$  1.03e-03
&  0.137 $\pm$  1.21e-03
&  0.127 $\pm$  1.30e-03
&  0.121 $\pm$  1.40e-03
&  0.118 $\pm$  1.41e-03
 \\\midrule
  \multirow{2}{*}{PBM (m=256)} 

&  0.893
&  0.904
&  0.904
&  0.900
&  0.897
&  0.898
&  0.897

  \\
&  0.779 $\pm$  2.12e-04
&  0.227 $\pm$  7.40e-04
&  0.160 $\pm$  1.03e-03
&  0.138 $\pm$  1.20e-03
&  0.127 $\pm$  1.31e-03
&  0.122 $\pm$  1.36e-03
&  0.118 $\pm$  1.40e-03
 \\\midrule
  
  \multirow{2}{*}{PBM (m=1024)} 
&  0.896
&  0.900
&  0.905
&  0.901
&  0.900
&  0.904
&  0.895

  \\

&  0.779 $\pm$  2.13e-04
&  0.227 $\pm$  7.36e-04
&  0.160 $\pm$  1.03e-03
&  0.137 $\pm$  1.19e-03
&  0.127 $\pm$  1.29e-03
&  0.121 $\pm$  1.36e-03
&  0.118 $\pm$  1.40e-03

\\\midrule

  \multirow{2}{*}{PBM (m=2048)} 
&  0.898
&  0.897
&  0.902
&  0.901
&  0.903
&  0.900
&  0.899\\

&  0.779 $\pm$  2.11e-04
&  0.227 $\pm$  7.30e-04
&  0.160 $\pm$  1.03e-03
&  0.137 $\pm$  1.20e-03
&  0.127 $\pm$  1.30e-03
&  0.121 $\pm$  1.40e-03
&  0.118 $\pm$  1.41e-03\\
  \bottomrule
\end{tabular}
\end{table}
}

{\small
\begin{table}[H]
\centering
\caption{Average width and coverage of $90\%$-confidence intervals for PATE. Constant treatment effect with $n=10^3$.}
\vspace{0.1 cm}
\label{tbl:exp_PATE_n_1000_constant}
\tiny
\setlength{\tabcolsep}{4.0pt}
\begin{tabular}{@{}cccccccc@{}}
\toprule

\multirow{2}{*}{Non-private} 

& \multicolumn{7}{c}{0.901}\\
& \multicolumn{7}{c}{0.002 $\pm$  3.24e-05}
\\ \midrule
\multicolumn{1}{c}{} &
  $\varepsilon = 0.1$ &
  $\varepsilon = 0.4$ &
  $\varepsilon = 0.7$ &
  $\varepsilon = 1.0$ &
  $\varepsilon = 1.3$ &
  $\varepsilon = 1.6$ &
  $\varepsilon = 1.9$ \\ \midrule
\multirow{2}{*}{Central Gaussian} 

&  0.905
&  0.895
&  0.899
&  0.902
&  0.904
&  0.899
&  0.899

\\

&  0.771 $\pm$  1.24e-07
&  0.199 $\pm$  4.85e-07
&  0.118 $\pm$  8.20e-07
&  0.084 $\pm$  1.16e-06
&  0.066 $\pm$  1.47e-06
&  0.055 $\pm$  1.78e-06
&  0.047 $\pm$  2.07e-06
 \\\midrule
  \multirow{2}{*}{PBM (m=256)} 
&  0.902
&  0.900
&  0.900
&  0.903
&  0.906
&  0.900
&  0.903 
  \\

&  0.772 $\pm$  1.25e-07
&  0.200 $\pm$  4.84e-07
&  0.119 $\pm$  8.15e-07
&  0.085 $\pm$  1.15e-06
&  0.067 $\pm$  1.43e-06
&  0.056 $\pm$  1.72e-06
&  0.048 $\pm$  2.02e-06

 \\\midrule
  
  \multirow{2}{*}{PBM (m=1024)}

&  0.900
&  0.897
&  0.902
&  0.900
&  0.904
&  0.898
&  0.896

  \\

&  0.772 $\pm$  1.26e-07
&  0.199 $\pm$  4.85e-07
&  0.118 $\pm$  8.28e-07
&  0.085 $\pm$  1.17e-06
&  0.066 $\pm$  1.46e-06
&  0.055 $\pm$  1.77e-06
&  0.047 $\pm$  2.05e-06

\\\midrule

  \multirow{2}{*}{PBM (m=2048)} 

&  0.897
&  0.902
&  0.901
&  0.901
&  0.899
&  0.902
&  0.898
\\

&  0.772 $\pm$  1.24e-07
&  0.199 $\pm$  4.85e-07
&  0.118 $\pm$  8.19e-07
&  0.084 $\pm$  1.16e-06
&  0.066 $\pm$  1.47e-06
&  0.055 $\pm$  1.77e-06
&  0.047 $\pm$  2.06e-06
\\
  \bottomrule
\end{tabular}
\end{table}
}

\subsection{Constant treatment effect with larger $n$}
Finally, in the last set of experiments, we consider a larger sample size with Gaussian outcomes. We use the same set of parameters as in Section~\ref{subsec:exp_gaussian}, except that $n_t = n_c = 10^4$. From Table~\ref{tbl:exp_SATE_n_1000_constant} and Table~\ref{tbl:exp_PATE_n_1000_constant}, we observe that when the privacy budget is large enough $\varepsilon > 1$, the CIs for both PBM and central Gaussian are very closed to the non-private one, indicating that the error is dominated by the sampling noise instead of the DP noise. Therefore, when $n$ is large enough (depending on the sample variance), we can achieve DP with negligible effect on the utility.
{\small
\begin{table}[H]
\centering
\caption{Average width and coverage of $90\%$-confidence intervals for SATE. Constant treatment effect with $n=10^4$.}
\vspace{0.1 cm}
\label{tb1:SATE_gauss_large_n}
\tiny
\setlength{\tabcolsep}{4.0pt}
\begin{tabular}{@{}cccccccc@{}}
\toprule

\multirow{2}{*}{Non-private} 

& \multicolumn{7}{c}{0.896}\\
& \multicolumn{7}{c}{ 0.034 $\pm$  1.51e-04}
\\ \midrule
\multicolumn{1}{c}{} &
  $\varepsilon = 0.1$ &
  $\varepsilon = 0.4$ &
  $\varepsilon = 0.7$ &
  $\varepsilon = 1.0$ &
  $\varepsilon = 1.3$ &
  $\varepsilon = 1.6$ &
  $\varepsilon = 1.9$ \\ \midrule
\multirow{2}{*}{Central Gaussian} 

&  0.905
&  0.903
&  0.896
&  0.903
&  0.899
&  0.899
&  0.903

\\

&  0.084 $\pm$  6.25e-05
&  0.040 $\pm$  1.32e-04
&  0.036 $\pm$  1.45e-04
&  0.035 $\pm$  1.49e-04
&  0.035 $\pm$  1.49e-04
&  0.035 $\pm$  1.52e-04
&  0.035 $\pm$  1.50e-04

 \\\midrule
  \multirow{2}{*}{PBM (m=256)} 

&  0.905
&  0.906
&  0.897
&  0.902
&  0.897
&  0.896
&  0.905

  \\

&  0.084 $\pm$  6.15e-05
&  0.040 $\pm$  1.32e-04
&  0.036 $\pm$  1.42e-04
&  0.036 $\pm$  1.48e-04
&  0.036 $\pm$  1.47e-04
&  0.036 $\pm$  1.46e-04
&  0.036 $\pm$  1.47e-04

 \\\midrule
  
  \multirow{2}{*}{PBM (m=1024)} 
&  0.899
&  0.897
&  0.902
&  0.902
&  0.898
&  0.903
&  0.900

  \\

&  0.085 $\pm$  6.16e-05
&  0.040 $\pm$  1.32e-04
&  0.036 $\pm$  1.45e-04
&  0.035 $\pm$  1.48e-04
&  0.035 $\pm$  1.49e-04
&  0.035 $\pm$  1.51e-04
&  0.035 $\pm$  1.52e-04

\\\midrule

  \multirow{2}{*}{PBM (m=2048)} 
&  0.903
&  0.903
&  0.899
&  0.898
&  0.906
&  0.898
&  0.901
\\

&  0.085 $\pm$  6.22e-05
&  0.040 $\pm$  1.31e-04
&  0.036 $\pm$  1.45e-04
&  0.035 $\pm$  1.49e-04
&  0.035 $\pm$  1.49e-04
&  0.035 $\pm$  1.51e-04
&  0.035 $\pm$  1.50e-04
\\
  \bottomrule
\end{tabular}
\end{table}
}

{\small
\begin{table}[H]
\centering
\caption{Average width and coverage of $90\%$-confidence intervals for PATE. Constant treatment effect with $n=10^4$.}
\vspace{0.1 cm}
\label{tb1:PATE_gauss_large_n}
\tiny
\setlength{\tabcolsep}{4.0pt}
\begin{tabular}{@{}cccccccc@{}}
\toprule

\multirow{2}{*}{Non-private} 

& \multicolumn{7}{c}{0.904}\\
& \multicolumn{7}{c}{ 0.034 $\pm$  1.53e-04}
\\ \midrule
\multicolumn{1}{c}{} &
  $\varepsilon = 0.1$ &
  $\varepsilon = 0.4$ &
  $\varepsilon = 0.7$ &
  $\varepsilon = 1.0$ &
  $\varepsilon = 1.3$ &
  $\varepsilon = 1.6$ &
  $\varepsilon = 1.9$ \\ \midrule
\multirow{2}{*}{Central Gaussian} 

&  0.904
&  0.899
&  0.903
&  0.907
&  0.900
&  0.900
&  0.900

\\

&  0.084 $\pm$  6.23e-05
&  0.040 $\pm$  1.33e-04
&  0.036 $\pm$  1.42e-04
&  0.035 $\pm$  1.50e-04
&  0.035 $\pm$  1.51e-04
&  0.035 $\pm$  1.51e-04
&  0.035 $\pm$  1.51e-04
 
 \\\midrule
  \multirow{2}{*}{PBM (m=256)} 

&  0.903
&  0.897
&  0.907
&  0.911
&  0.901
&  0.900
&  0.899

  \\

&  0.084 $\pm$  6.17e-05
&  0.040 $\pm$  1.30e-04
&  0.036 $\pm$  1.43e-04
&  0.036 $\pm$  1.49e-04
&  0.036 $\pm$  1.46e-04
&  0.036 $\pm$  1.45e-04
&  0.036 $\pm$  1.47e-04

 \\\midrule
  
  \multirow{2}{*}{PBM (m=1024)} 

&  0.899
&  0.898
&  0.905
&  0.903
&  0.904
&  0.901
&  0.896

  \\

&  0.085 $\pm$  6.15e-05
&  0.040 $\pm$  1.33e-04
&  0.036 $\pm$  1.46e-04
&  0.035 $\pm$  1.48e-04
&  0.035 $\pm$  1.50e-04
&  0.035 $\pm$  1.52e-04
&  0.035 $\pm$  1.50e-04

\\\midrule

  \multirow{2}{*}{PBM (m=2048)} 

&  0.905
&  0.900
&  0.901
&  0.903
&  0.903
&  0.896
&  0.901

\\

&  0.085 $\pm$  6.21e-05
&  0.040 $\pm$  1.32e-04
&  0.036 $\pm$  1.42e-04
&  0.035 $\pm$  1.50e-04
&  0.035 $\pm$  1.51e-04
&  0.035 $\pm$  1.50e-04
&  0.035 $\pm$  1.51e-04

\\
  \bottomrule
\end{tabular}
\end{table}
}

\section{Omitted Proofs}
\subsection{Proof of Theorem~\ref{thm:SATE_ddp}}\label{sec:additional_proofs}

We follow the standard analysis of the difference-in-mean estimator and incorporate the DP noise. To begin with, we analyze the unprivatized estimator. Let $\hat{\nu}_t \eqDef \frac{1}{n_t} \sum_{i}T_i y_i(0)$ and $\hat{\nu}_c \eqDef \frac{1}{n_c}\sum_i (1-T_i)y_i(1)$ be the unprivatized means of the test and control groups. In addition, let $s^2_c \eqDef \frac{1}{n-1}\sum_{i} (y_i(0) - \bar{y}(0))^2$ and $s^2_t \eqDef \frac{1}{n-1}\sum_{i} (y_i(1) -\bar{y}(1))^2$ be the sample variances; let $s_{tc} \eqDef \frac{1}{n-1}\sum_{i}(y_i(0)-\bar{y}(0))(y_i(1)- \bar{y}(1))$ be the sample covariance.  Then, the variance of the (unprivatized) difference-in-mean estimator can be computed as
\begin{align*}
    \Var\lp \hat{\nu}_t - \hat{\nu}_c \mv \mb{y} \rp = 
    \frac{\sigma^2_{s}}{n} \eqDef \frac{1}{n}\lp \frac{n_c}{n_t} s^2_t + \frac{n_t}{n_c} s^2_c + s_{tc} \rp.
\end{align*}

The finite-sample central limit theorem \citep{hajek1961some} (see also \citet{li2017general, li2018asymptotic}) suggests that 
\[\sqrt{n}\lp (\hat{\nu}_t - \hat{\nu}_c) - \Delta_s \rp \dra N(0, \sigma^2_{s}). \]

When there exists DP noise, we have, conditioned on $\mb{y}$ and $T_i$,
\begin{align*}
    \sqrt{n}\lp (\hat{\mu}_c - \hat{\mu}_t) - (\hat{\nu}_t - \hat{\nu}_c)\rp
    \dra N\lp 0, \sigma^2_\msf{pr}(n_c, n_t, \varepsilon)\rp,
\end{align*}
where $\sigma^2_\msf{pr}(n_c, n_t, \varepsilon) \eqDef \frac{n}{n_c}\sigma^2_1(n_c, \varepsilon) + \frac{n}{n_t}\sigma^2_1(n_t, \varepsilon)$ and the convergence is due to the (classical) central limit theorem and Assumption~\ref{assumption:variance}. Since the DP noise is independent with $T_i$, we conclude 
\begin{align*}
    \sqrt{n}\lp (\hat{\mu}_c - \hat{\mu}_t) - \Delta_s\rp
    \dra N\lp 0, \sigma^2_\msf{pr}(n_c, n_t, \varepsilon) + \sigma^2_{s})\rp,
\end{align*}

Finally, since $\hat{\sigma}_\msf{s}^2$ defined in Algorithm~\ref{alg:SATE_ddp_general} is a high probability upper bound on $\sigma^2_\msf{s}$ from our assumptions, i.e.,
\[ \lim_{n\ra \infty} \Pr\lbp\hat{\sigma}_\msf{s}^2 \geq \sigma^2_\msf{s}  \rbp = 1, \]
by Slutsky's theorem $\hat{\Delta}_\msf{s} \pm z_{1-\alpha/2}\cdot  \lp \hat{\sigma}_\msf{s} + \sigma_{pr} \rp$ gives an $(1-\alpha)$-CI asymptotically.

Next, we prove the coverage guarantee for estimating PATE. Observe that the conditional variance of the (unprivatized) difference-in-mean estimator, given the samples $y_i(0) \diid P_0$ and $y_i(1) \diid P_1$, can be computed as
\begin{align*}
    \Var\lp \hat{\nu}_t - \hat{\nu}_c \mv \bm{y} \rp = \frac{1}{n}\lp \frac{n_c}{n_t} s^2_t + \frac{n_t}{n_c} s^2_c + s_{tc} \rp.
\end{align*}

Therefore, the unconditional variance is 
\begin{align*}
    \E\lb \Var\lp \hat{\nu}_t - \hat{\nu}_c \mv \bm{y} \rp\rb + \Var\lp \E\lb \hat{\nu}_t - \hat{\nu}_c \mv \bm{y} \rb  \rp & =  \frac{1}{n}\lp \frac{n_c}{n_t} s^2_t + \frac{n_t}{n_c} s^2_c + 2s_{tc} \rp + \frac{1}{n}\lp s^2_t + s^2_c - 2s_{tc} \rp \\
    & = \frac{s^2_t}{n_t} + \frac{s^2_c}{n_c}.
\end{align*}

As a result, $\hat{\sigma}_\msf{p}$ in Algorithm~\ref{alg:SATE_ddp_general} is a consistent estimator of the variance of the unprivatized estimator.

With the presence of DP noise, we follow the same analysis as SATE and add a calibration term $\sigma^2_\msf{pr}(n_c, n_t, \varepsilon)$. By the central limit theorem, the proof is complete. \qedwhite

\subsection{Proof of Theorem~\ref{thm:pate_CIs}}\label{appendix:proof_of_pate_berstein}

We first present the full version of the theorem with higher-order terms and constants.
\begin{theorem}[Detailed version of Theorem~\ref{thm:pate_CIs}]\label{thm:pate_CIs_detailed}
    Let $\mcal{M}_1$ and $\mcal{M}_2$ be PBM (Algorithm~\ref{alg:scalar_pbm}) with parameter $(m_1, \theta_1)$ and $(m_2, \theta_2)$. Let $\hat{\sigma}_\msf{p}$ be defined as in Algorithm~\ref{alg:SATE_ddp_general}. Then under Assumption~\ref{assumption:bdd_x}, it holds that
    \begin{align*}
        \Pr\lbp \Delta_\msf{p} \in \lb \hat{\Delta}_\msf{p} - \lp \sqrt{2\hat{\sigma}^2_\msf{p} \log(2.01/\delta)} + \gamma\rp , \hat{\Delta}_\msf{p} + \lp \sqrt{2\hat{\sigma}^2_\msf{p} \log(2.01/\delta)} + \gamma\rp  \rb \rbp \geq 1-\delta,
    \end{align*}
    where 
    \begin{align*}
    &\gamma = \frac{56R\log(1200/\delta)}{3(n-1)} + \sqrt{\frac{R^2}{2m_1n\theta_1^2}\log\lp \frac{1200}{\delta} \rp} + \sqrt{\frac{4\log(2.01/\delta_1)}{n}} \nonumber\\
    & \quad \quad \cdot \min\lp
     \sqrt[4]{\log\lp \frac{1200}{\delta}\rp\frac{R^4}{4m_2n\theta_2^2}} + \sqrt[4]{\log\lp \frac{1200}{\delta} \rp \frac{R^2}{2m_1n\theta_1^2}}, 
    \quad \frac{\sqrt{\log\lp \frac{1200}{\delta}\rp\frac{R^4}{4m_2n\theta_2^2}} + \sqrt{\log\lp \frac{1200}{\delta} \rp \frac{R^2}{2m_1n\theta_1^2}}}{\sqrt{\hat{s}^2_t + \hat{s}^2_c} }
    \rp.
\end{align*}
Note that when $\varepsilon_1(\alpha) = C_1 \frac{\alpha m_1 \theta_1^2}{n}$ and $\varepsilon_2(\alpha) = C_2 \frac{\alpha m_2 \theta_2^2}{n}$ are constants,  $\gamma = O(1/n)$.
\end{theorem}

\textbf{Proof.}
Before entering the main proof, we will make use of the following (slightly adapted) empirical Berstein inequality:
\begin{lemma}[Theorem~11, \citet{maurer2009empirical}]\label{lemma:empirical_bernstein_id}
    Let $\mb{X} = (X_1, ..., X_n)$ be a vector of independent random variables with values in $[-R, R]$. Let $\delta > 0$. Then for any $\delta_1, \delta_2 > 0$ and $\delta_1 + \delta_2 = \delta$, it implies
    \begin{equation}\label{eq:empirical_berstein}
        \Pr\lbp  \lba \frac{1}{n}\sum_i X_i - \frac{1}{n}\sum_i\E[X_i] \rba \leq \sqrt{\frac{2s^2\lp \mb{X} \rp \log\lp \frac{2}{\delta_1} \rp}{n}} + \frac{14R\log\lp \frac{2}{\delta_2} \rp}{3(n-1)}  \rbp \geq 1-\delta,
    \end{equation}
    where $s^2(\mb{X}) \eqDef \frac{1}{n(n-1)}\sum_{i\neq j}\lp X_i - X_j \rp^2$ denotes the sample variance.
\end{lemma}
Now, we apply the above lemma in our PATE estimation task. Under the PATE setting with $n_c = n_t = n/2$, it holds that $X_1,...,X_{n/2} \diid P_t$ and $X_{n/2+1},...,X_n \diid P_c$. As a result, Lemma~\ref{lemma:empirical_bernstein_id} yields that, with probability $1-\delta$,
\begin{align}\label{eq:pate_berstein}
    \lba \frac{2}{n}\sum_{i\in[n/2]} \underbrace{\lp X_i - X_{i+n/2}\rp}_{\eqDef W_i} - \Delta_\msf{p} \rba 
   & \leq \sqrt{\frac{4 s^2\lp\bm{W}\rp\log(1/\delta_1)}{n}} + \frac{56R\log(2/\delta_2)}{3(n-1)} \nonumber\\
    & =  \sqrt{\frac{4 \lp s^2_t + s^2_c\rp \log(1/\delta_1)}{n}} + \frac{56R\log(2/\delta_2)}{3(n-1)},
\end{align}
where the first inequality holds since $W_i \in [-2R, 2R]$ and the second equality holds since 
$$ s^2(\bm{W}) = s^2\lp X_1,...,X_{n/2} \rp + s^2\lp X_{n/2+1},..., X_{n} \rp = s^2_t + s^2_c $$
by definition.

Next, it suffices to combine with the concentration bounds on $\hat{\Delta}_\msf{p} \eqDef \hat{\mu}_t - \hat{\mu}_c$ and $\hat{s}^2_t$ and $\hat{s}^2_c$ (recall that these are the private estimates of sample means and variance from PBM).

\paragraph{Concentration of private sample mean.} To this end, observe that
\begin{align*}
    \hat{\mu}_t - \hat{\mu}_c = \frac{2R}{nm_1\theta}\sum_{i=1}^{n}\lp Z_i - \frac{m_1}{2} \rp,
\end{align*}
where $Z_i \sim \msf{Binom}\lp m_1, \frac{1}{2}-\frac{\theta_1 X_{i+n/2}}{R} \rp$ for $i \in [n/2]$ and $Z_i \sim \msf{Binom}\lp m_1, \frac{1}{2}+\frac{\theta_1 X_{i+n/2}}{R} \rp$ for $i \in [n/2+1:n]$.
Conditioning on $X_i$'s and applying Hoeffding's inequality yield
\begin{align}\label{eq:pate_mu_bdd}
    &\Pr\lbp \lba \lp \hat{\mu}_t - \hat{\mu}_c\rp - \frac{2}{n}\sum_{i=1}^{n/2} \lp X_{i} - X_{i+n/2}\rp \rba \geq  \sqrt{\log\lp \frac{2}{\delta_\mu} \rp \frac{R^2}{2m_1n\theta_1^2}} \rbp \nonumber\\
    &= \Pr\lbp \lba \underbrace{\sum_{i=1}^{n}Z_i -   \sum_{i=1}^{n/2} \lp \frac{m_1\theta_1}{R}\lp X_{i} - X_{i+n/2}\rp - \frac{m_1}{2}\rp }_{\text{ sum of } m_1n/2 \text{ independent zero-mean bounded variables.}}\rba \geq \sqrt{\log\lp \frac{2}{\delta_\mu} \rp \frac{m_1n}{2}} \rbp \nonumber \\
    &\leq \delta_\mu.
\end{align}

\paragraph{Concentration of private sample variance.} Next, we construct the estimator of the sample variance from PBM:
$$ \hat{s}^2_t = \frac{1}{n-1}\sum_{i=1}^{n/2} \lp \frac{R^2}{2m_2\theta_2}Z_i'- \frac{R^2}{4\theta_2} + \frac{R^2}{2}\rp - \frac{n}{n-1}\hat{\mu}_t^2,  $$
where $Z_i' \sim \msf{Binom}\lp m_2, 2\theta\lp\frac{X_i^2}{R^2}-\frac{1}{2}\rp + \frac{1}{2} \rp$. We construct $\hat{s}_c$ in the same way.
Notice that the above $\hat{s}^2_t$ is constructed such that $\E\lb \hat{s}^2_t \mv \bm{X} \rb = s^2_t$.

Since the sample variance estimator $\hat{s}^2_t$ and $\hat{s}^2_c$ are privatized by PBM, it is possible to obtain negative values, so we will replace them by its positive part, i.e., $\hat{s}^{2+}_t \eqDef \max\lp \hat{s}^{2}_t, 0 \rp$ and $\hat{s}^{2+}_c \eqDef \max\lp \hat{s}^{2}_c, 0 \rp$. For notational convenience, we abuse notation and let $\hat{s}^2_t$ and $\hat{s}^2_c$ be the positive parts so that $\hat{s}^2_t, \hat{s}^2_c \geq 0$ always holds.

Since $\hat{s}^2_t$ is obtained by first estimating the second moment of samples $\sum_i X_i^2$ and then subtract the sample mean $n\hat{\mu}_t^2$, it holds that, conditioning on $X_i$'s and the event 
$$\lbp \hat{\Delta}_\msf{p} - \frac{2}{n}\sum_{i=1}^{n/2}\lp X_i - X_{i+n/2} \rp < \sqrt{\log\lp \frac{2}{\delta_\mu} \rp \frac{R^2}{2m_1n\theta_1^2}} \rbp, $$ 
we have
\begin{align}\label{eq:pate_var_bdd}
    \Pr\lbp \lba \lp \hat{s}^2_t + \hat{s}^2_c \rp - \lp s_t^2 + s_c^2 \rp \rba \geq \sqrt{\log\lp \frac{2}{\delta_s}\rp\frac{R^4}{4m_2n\theta_2^2}} + \sqrt{\log\lp \frac{2}{\delta_\mu} \rp \frac{R^2}{2m_1n\theta_1^2}}\rbp \leq \delta_s.
\end{align}
This implies that with probability at least $1-\delta_s$, both of the following events hold:
\begin{itemize}
    \item \begin{align*}
        \sqrt{\hat{s}^2_t + \hat{s}^2_c}  
        & \leq \sqrt{\lp s_t^2 + s_c^2 \rp + \sqrt{\log\lp \frac{2}{\delta_s}\rp\frac{R^4}{4m_2n\theta_2^2}} + \sqrt{\log\lp \frac{2}{\delta_\mu} \rp \frac{R^2}{2m_1n\theta_1^2}}} \\
        & \leq \sqrt{ s_t^2 + s_c^2 } + \sqrt[4]{\log\lp \frac{2}{\delta_s}\rp\frac{R^4}{4m_2n\theta_2^2}} + \sqrt[4]{\log\lp \frac{2}{\delta_\mu} \rp \frac{R^2}{2m_1n\theta_1^2}}.
    \end{align*}  

    \item 
    \begin{align*}
        \sqrt{\hat{s}^2_t + \hat{s}^2_c}  
        & \leq \sqrt{s_t^2 + s_c^2 } + \frac{\sqrt{\log\lp \frac{2}{\delta_s}\rp\frac{R^4}{4m_2n\theta_2^2}} + \sqrt{\log\lp \frac{2}{\delta_\mu} \rp \frac{R^2}{2m_1n\theta_1^2}}}{\sqrt{\hat{s}^2_t + \hat{s}^2_c} + \sqrt{s_t^2 + s_c^2 }} \\
        &  \leq \sqrt{ s_t^2 + s_c^2 } + \frac{\sqrt{\log\lp \frac{2}{\delta_s}\rp\frac{R^4}{4m_2n\theta_2^2}} + \sqrt{\log\lp \frac{2}{\delta_\mu} \rp \frac{R^2}{2m_1n\theta_1^2}}}{\sqrt{\hat{s}^2_t + \hat{s}^2_c} }.
    \end{align*}  
\end{itemize}
Therefore, we arrive at the following bound on the private sample variance:
\begin{align}\label{eq:pate_variance_bdd}
    \Pr\bigg( \sqrt{\hat{s}^2_t + \hat{s}^2_c} \geq \sqrt{ s_t^2 + s_c^2 } +
    \min\bigg( 
    & \sqrt[4]{\log\lp \frac{2}{\delta_s}\rp\frac{R^4}{4m_2n\theta_2^2}} + \sqrt[4]{\log\lp \frac{2}{\delta_\mu} \rp \frac{R^2}{2m_1n\theta_1^2}}, \nonumber\\
    & \quad \frac{\sqrt{\log\lp \frac{2}{\delta_s}\rp\frac{R^4}{4m_2n\theta_2^2}} + \sqrt{\log\lp \frac{2}{\delta_\mu} \rp \frac{R^2}{2m_1n\theta_1^2}}}{\sqrt{\hat{s}^2_t + \hat{s}^2_c} }
    \bigg) \bigg) \leq \delta_s.
\end{align}

\paragraph{Putting things together.}
Finally, by plugging \eqref{eq:pate_mu_bdd} and \eqref{eq:pate_variance_bdd} into \eqref{eq:pate_berstein}, we obtain that, with probability at least $1-\delta_1 -\delta_2 - \delta_\mu - \delta_s$,
\begin{align}
    \lba \hat{\Delta}_{\msf{p}} - \Delta_{\msf{p}}\rba \leq \sqrt{\frac{4\lp \hat{s}^2_t + \hat{s}^2_c \rp \log(2/\delta_1)}{n}} + \gamma =  \sqrt{2\hat{\sigma}^2_\msf{p} \log(2/\delta_1)} + \gamma,
\end{align}
where $\gamma = o(1/\sqrt{n})$ and takes the following explicit expression:
\begin{align*}
    &\gamma = \frac{56R\log(2/\delta_2)}{3(n-1)} + \sqrt{\frac{R^2}{2m_1n\theta_1^2}\log\lp \frac{2}{\delta_\mu} \rp}
    +\sqrt{\frac{4\log(2/\delta_1)}{n}}\cdot \\
    &\min\lp 
     \sqrt[4]{\log\lp \frac{2}{\delta_s}\rp\frac{R^4}{4m_2n\theta_2^2}} + \sqrt[4]{\log\lp \frac{2}{\delta_\mu} \rp \frac{R^2}{2m_1n\theta_1^2}}, \nonumber
     \frac{\sqrt{\log\lp \frac{2}{\delta_s}\rp\frac{R^4}{4m_2n\theta_2^2}} + \sqrt{\log\lp \frac{2}{\delta_\mu} \rp \frac{R^2}{2m_1n\theta_1^2}}}{\sqrt{\hat{s}^2_t + \hat{s}^2_c} }
    \rp.
\end{align*}
Finally, we can pick $\delta_1 = 0.995\delta$ and $\delta_2 = \delta_\mu = \delta_s = \frac{\delta}{600}$, which yields
\begin{align}
    \lba \hat{\Delta}_{\msf{p}} - \Delta_{\msf{p}}\rba \leq \sqrt{\frac{4\lp \hat{s}^2_t + \hat{s}^2_c \rp \log(2.01/\delta)}{n}} + \gamma =  \sqrt{2\hat{\sigma}^2_\msf{p} \log(2.01/\delta)} + \gamma,
\end{align}
and 
\begin{align*}
    &\gamma = \frac{56R\log(1200/\delta)}{3(n-1)} + \sqrt{\frac{R^2}{2m_1n\theta_1^2}\log\lp \frac{1200}{\delta} \rp} + \sqrt{\frac{4\log(2.01/\delta_1)}{n}} \nonumber\\
    & \quad \quad \cdot \min\lp
     \sqrt[4]{\log\lp \frac{1200}{\delta}\rp\frac{R^4}{4m_2n\theta_2^2}} + \sqrt[4]{\log\lp \frac{1200}{\delta} \rp \frac{R^2}{2m_1n\theta_1^2}}, 
    \quad \frac{\sqrt{\log\lp \frac{1200}{\delta}\rp\frac{R^4}{4m_2n\theta_2^2}} + \sqrt{\log\lp \frac{1200}{\delta} \rp \frac{R^2}{2m_1n\theta_1^2}}}{\sqrt{\hat{s}^2_t + \hat{s}^2_c} }
    \rp.
\end{align*}
\qedwhite
\subsection{Proof of Proposition~\ref{prop:pate_split}}
Since $\mcal{M}_1$ and $\mcal{M}_2$ satisfy $(\alpha, \varepsilon_1(\alpha))$ and $(\alpha, \varepsilon_2(\alpha))$ R\'enyi DP, it holds that $m_1\theta_1^2 \leq \frac{Cn\varepsilon_1(\alpha)}{\alpha}$ and $m_2\theta_2^2 \leq \frac{Cn\varepsilon_2(\alpha)}{\alpha}$ for some universal constant $C$.
Plugging these into Theorem~\ref{thm:pate_CIs_detailed} yields the desired result.
\qedwhite

\subsection{Proof of Theorem~\ref{thm:sate_CIs}}\label{appendix:proof_sate_berstein}
The proof follows from the same step as in the proof of Theorem~\ref{thm:pate_CIs}, except for replacing the empirical Berstein inequality with the following finite sample (i.e., without-replacement) Berstein inequality:
\begin{lemma}[Proposition~1.4 of \citet{bardenet2015concentration}]\label{lemma:berstein_wr}
    Let $\mcal{X} = \{ x_1, x_2,...,x_n\}$ be a finite set of $N$ points. Let 
    $$ a \eqDef \min_{i\in[n]}x_i, \text{ and } b \eqDef \max_{i\in[n]} x_i;$$
    $$ \mu \eqDef \frac{1}{n}\sum_{i\in[n]} x_i \text{ and } \sigma^2 \eqDef \frac{1}{n}\sum_{i\in[n]}(x
_i - \mu)^2. $$ Let $X_1, X_2,...,X_{n/2}$ denote a random sample drawn without replacement from $\mcal{X}$. Then, for all $\varepsilon>0$,
\begin{equation}
    \Pr\lp \lba \frac{1}{n}\sum_{i\in[n]}X_i - \mu \rba \geq \varepsilon \rp \leq 2\exp\lp -\frac{n\varepsilon^2}{2\sigma^2 + (2/3)(b-a)\varepsilon} \rp.
\end{equation}
\end{lemma}
Notice that Lemma~\ref{lemma:berstein_wr} implies Lemma~\ref{lemma:empirical_bernstein_id} with $\E[X_i]$ being replaced by $\mu$ and $s^2\lp \mb{X} \rp$ replaced by $\sigma\lp x_1,...,x_n \rp$. As a result, we only need to apply concentration inequalities on the private estimate of $\hat{\mu}_{\msf{pr}}$ and $\hat{\sigma}^2_{\msf{s}}$, which follows from the proof of Theorem~\ref{thm:pate_CIs}.

\qedwhite

\end{document}